\documentclass[a4paper,11pt]{article}

\pdfoutput=1

\usepackage{jheppub}
\usepackage{color}
\usepackage{graphicx}                   
\usepackage{amsmath}
\usepackage{amssymb}
\usepackage{amsfonts}
\usepackage{xspace}
\usepackage{verbatim}

\usepackage[small]{subfigure}
\usepackage{slashed}
\usepackage{mdwlist}
\usepackage[normalem]{ulem}
\usepackage[usenames,dvipsnames,svgnames,table]{xcolor}

\newcommand{\powR}{2}
\newcommand{\Rir}{R_{ir}}
\newcommand{\Ruv}{R_{uv}}

\newcommand{\zh}{z_h}
\newcommand{\chidof}{$\chi^2/d.o.f.\,$}
\newcommand{\ffa}{d_{\sphericalangle}}
\newcommand{\fxseca}{D_{\sphericalangle}}
\newcommand{\ffh}{d}
\newcommand{\fxsech}{D}

\newcommand{\barffa}{\bar{d}_{\sphericalangle}}
\newcommand{\barfxseca}{\bar{D}_{\sphericalangle}}
\newcommand{\barffh}{\bar{d}}
\newcommand{\barfxsech}{\bar{D}}
\allowdisplaybreaks[2]

  \newcount\hour \newcount\minute
  \hour=\time \divide \hour by 60 \minute=\time
  \newcommand{\todaytime}{\today \ -- \number\hour :\ifnum \minute<10 0\fi\number\minute}

\title{The Fragmentation Spectrum from Space-Time Reciprocity}

\author[a]{Duff Neill}

\affiliation[a]{Theoretical Division, MS B283, Los Alamos National Laboratory, Los Alamos, NM 87545, USA}
                                         
\emailAdd{duff.neill@gmail.com}

\abstract{Analyzing the single inclusive annihilation spectrum of charged hadrons in $e^+e^-$ collisions, I confront the hadronization hypothesis of local parton-hadron duality with a systematic resummation of the dependence on the small energy fraction. This resummation is based on the reciprocity between time-like and space-like splitting processes in $4-2\epsilon$-dimensions, which I extend to resum all the soft terms of the cross-section for inclusive jet production. Under the local-parton-hadron duality hypothesis, the resulting distribution of jets essentially determines the spectrum of hadrons as the jet radius goes to zero. Thus I take the resummed perturbative jet function as the non-perturbative fragmentation function with an effective infra-red coupling. I find excellent agreement with data, and comment on the mixed leading log approximation previously used to justify local parton-hadron duality.}

\begin{document}
\maketitle
\section{Introduction}

Local Parton-Hadron Duality (LPHD) posits that the momentum spectrum of partons, when calculated perturbatively, essentially determines the momentum spectrum of the resulting hadrons \cite{Azimov:1984np}, so long as there is an underlying hard process at a scale well-above confinement. LPHD is consistent with the much more rigorous statement of factorization, which states that physics at different scales is not entangled beyond the anomalous dimensions of scaling quantities. Within QCD factorization, factorized sectors can be treated independently up to power corrections, and up to consistency conditions on their divergences. In a factorization theorem, any sector sensitive to scales where the running coupling is perturbative can be legitimately calculated in perturbation theory. In contrast, when taken in its strong form the LPHD hypothesis goes well beyond the statement of factorization, since one wishes to apply perturbation theory as close to the non-perturbative regime as possible. This is to say that the perturbative results for infra-red non-perturbative functions are to be the model for such functions. This can be illustrated with the classic observable of fragmentation: the total energy $Q$ of the collision is broken up into various fractions, which are distributed amongst the asymptotic remnants of the hard interaction, labeled by $h$:
\begin{align}
\zh=2\frac{Q\cdot p_h}{Q^2}=2E_h/Q\text{ with }E_h\text{ being the energy of remnant } h\,.
\end{align}  
Within factorization, one constructs the fragmentation spectrum as the convolution of a non-perturbative boundary condition, the fragmentation function, with a perturbatively calculable renormalization group evolution, and a perturbatively calculable coefficient function. However, within LPHD, up to questions about the scale of confinement which ends all colored interactions between fields, a purely perturbative calculation should also determine the \emph{shape} of the fragmentation function. This is because the hadronization process itself is hypothesized to be unable to significantly change the pre-hadronization momentum distribution. Historically, this hypothesis when combined with a resummation of soft physics in the leading order Dokshitzer-Gribov-Lipatov-Alterelli-Parisi (DGLAP) equations led to the so-called ``mixed leading log'' approximation (MLLA), for review, ref. \cite{Khoze:1996dn}. 

From the basic equations for the MLLA as applied to fragmentation,\footnote{For a review of fragmentation, see ref. \cite{Metz:2016swz}.} two models may be derived that have seen phenomenological success: the distorted Gaussian approach of ref. \cite{Fong:1989qy,Fong:1990nt},\footnote{Global fits can be found in ref. \cite{Albino:2004yg} with the distorted Gaussian approximation to the MLLA, where the best fits require 5 parameters.} and the so-called limiting spectrum of ref. \cite{Dokshitzer:1991ej} (and references therein). Both are intended to describe the hadron fragmentation spectrum in the region where the bulk of hadrons are produced, as the energy fraction vanishes. Both approaches however have difficulties in describing the full small energy fraction region while consistently extrapolating between different center-of-mass energies for the $e^+e^-$ event.\footnote{While the limiting spectrum was used to describe OPAL data (ref. \cite{Akrawy:1990ha}) at the Large Electron-Positron (LEP) collider, and the TASSO and TOPAZ data, their fit does not extrapolate to the highest available energies. In ref. \cite{Perez-Ramos:2013eba}, the MLLA is extended to include higher order logs, which seems to improve the situation.} Moreover, it is unclear how exactly the MLLA relates to genuine perturbative QCD calculations of the $\overline{\text{MS}}$-splitting functions of QCD. While it is clear that the leading order DGLAP kernel acts as the leading order kernel for the MLLA equations, none-the-less, the MLLA derived anomalous dimension simply differs from the next-to-leading logarithmic $\overline{\text{MS}}$ anomalous dimension from small-$\zh$ resummed perturbation theory of refs. \cite{Vogt:2011jv,Kom:2012hd}. While this does not automatically invalidate the MLLA, one cannot simply match its results to known QCD perturbation theory beyond leading log.

Expanding upon a previous paper, ref. \cite{Neill:2020bwv}, I can construct a resummation of the soft terms in the cross-section that does not suffer from these limitations and derive a hadron fragmentation spectrum fully consistent with the standard perturbation theory results. I can precisely trace all quantities needed to define the hadron fragmentation spectrum to the \emph{space}-like DGLAP anomalous dimensions in $4-2\epsilon$-space-time dimensions. I do so by approaching the problem of hadrons as a problem of resumming inclusive jet production (\cite{Dasgupta:2014yra,Dasgupta:2016bnd,Kang:2016mcy,Dai:2016hzf}) when the radius of the jet approaches $0$, an idea initially explored in ref. \cite{Kang:2017mda}. Unlike the fragmentation function, the semi-inclusive jet function is perturbatively calculable for small but not extreme jet radii. However, one needs the jet function beyond the first few orders of perturbation theory, since DGLAP evolution drives all momentum fractions towards zero. Thus the jet function is dominated by its behavior at small energy fractions as the radius of the jet goes to zero. After I accomplish the resummation of the soft logarithms in the jet function, I can invoke the LPHD hypothesis to derive the model I need for the non-perturbative fragmentation function: the jet function evaluated with a frozen coupling scale in the infra-red. This is analogous to replacing the perturbative coupling with the effective coupling of ref. \cite{Dokshitzer:1995qm}, see also ref. \cite{Mattingly:1993ej}.

With this model of fragmentation, I can then fit the measured spectrum of charged hadrons with 3 parameters: the scale at which I freeze the coupling, the scale at which I end DGLAP evolution, and the basic normalization.  I will find an excellent quantitative description of data from the Large Electron Positron Collider (LEP) and the Stanford Linear Accelerator at the Z-pole for essentially all energy fractions below $\sim 0.5$. A good description is also obtained for older data at smaller center-of-mass energies, though hadron mass-corrections and issues of normalization limit the description of the extremely small-$\zh$ range.

Fits to the small-$\zh$ pion data have been made in ref. \cite{Anderle:2016czy} using a systematic resummation for the time-like splitting functions and the coefficient function derived in \cite{Albino:2011si,Vogt:2011jv,Kom:2012hd}, where the derived resummation for these quantities is equal to that obtained in ref. \cite{Neill:2020bwv}. But the corresponding resummation for the fragmentation function was not attempted. Rather, a generic model for the fragmentation function was used, with success at describing the data.

The organization of the paper is as follows. First, I extend the formalism of ref. \cite{Neill:2020bwv}, and show how to resum the semi-inclusive jet function/fragmentation function. Given the form of the resummation, one can expect large cancellations between the resummation of the coefficient function and the jet function, which naturally suggests an organization of resummed spectrum. I then compare this arrangement to the MLLA approximation. Finally, I give the non-perturbative model for fragmentation suggested by LPHD and the resummed perturbation theory, and compare to data. Then I conclude.

I note that all results in this paper equally apply to semi-inclusive jet production, and in what follows I will always call the low-scale semi-inclusive jet function simply the ``fragmentation function'' to avoid the somewhat cumbersome ``semi-inclusive jet function.'' Ultimately, the only difference between fragmenting to a jet and fragmenting to a hadron is the scale at which the evolution is stopped and the collinear function is evaluated, the functional form in ln$\zh$ being \emph{identical}.

\section{Resumming the fragmentation function}\label{sec:intro}
In this section I will explain how to resum the fragmentation function. 

I note the conventions for moment transforms as:
\begin{align}
\bar{f}(n)&=\int_{0}^{1}\frac{dx}{x}x^n\Big(xf(x)\Big)\,,  \\
xf(x)&=\int\displaylimits_{c-i\infty}^{c+i\infty}\frac{dn}{2\pi i} x^{-n}\bar{f}(n)\,.
\end{align}
The constant $c$ is chosen to place all singularities of $\bar{f}$ to the left of the contour. The soft region of $\zh\rightarrow 0$ corresponds to $n\rightarrow 0$. The log counting I adopt is $\alpha_s\sim n^2$, and the order of the pole in $n$ at $0$ corresponds to the number of logarithms in momentum space:
\begin{align}
\frac{1}{n^k}\leftrightarrow \frac{1}{x}\text{ln}^{k-1}x
\end{align}  
I start with the basic angular-ordered factorization posited in ref. \cite{Neill:2020bwv}, outline its connection to the standard formulation of the fragmentation spectrum in the factorization approach of perturbative QCD, focusing on the case of pure Yang-Mills theory. I generalize to full QCD in the following, as required to compare to data.

\subsection{Defining the resummed fragmentation function}
In the classic case of single-inclusive annihilation, one has a time-like hard momentum $Q$ injected via the off-shell decay of a color-singlet current, and the objective is to know how many of the final state particles each carry the fraction $\zh$ of $Q$. This number density of hadrons carrying fraction $\zh$ factorizes as (ref. \cite{Collins:1989gx}):
\begin{align}\label{eq:frag_factor_standard}
  \fxsech\Big(\zh,Q^2,\Lambda^2\Big)&=\int_{\zh}^{1}\frac{dz}{z}C\Big(z,Q^2,\mu^2\Big)\ffh\Big(\frac{\zh}{z},\mu^2,\Lambda^2\Big)+...\,,
\end{align}  
$C$ is the perturbative coefficient function describing the production of a hard parton with momentum fraction $z$. Subsequently that parton decays into hadrons, including $h$. This hard parton is coupled to the color-singlet current via fluctuations at the scale $Q$. The fragmentation function $\ffh$ describes how to convert this parton into hadrons, and is inherently non-perturbative, sensitive to the confinement scale $\Lambda$. Thus in perturbation theory $\fxsech$ is infra-red divergent.\footnote{In what follows, $\fxsech$ and $\ffh$ will always refer to the cross-section and fragmentation function in dimensional regularization, with or without infra-red divergences removed.}

When one calculates the cross-section $\fxsech$ in dimensional regularization to handle these infra-red divergences, then after the renormalization of ultra-violet divergences, one can take the fragmentation function $\ffh$ to absorb the infra-red divergences of the bare calculation of the number density, and thus acts as a counter-term for these IR divergences. Going to moment space, one has:
\begin{align}\label{eq:Factor_IR_Divergences}
  &\barfxsech\Big(n,Q^2,\Lambda^2\Big)=\text{exp}\Big(\int\displaylimits_{0}^{\alpha_s(\mu^2)}\frac{d\alpha}{\beta(\alpha,\epsilon)}\gamma^T(\alpha,n)\Big)\,\bar{C}\Big(n,Q^2,\mu^2\Big),\\
  &\text{ so }\barffh\Big(n,\mu^2,\Lambda^2\Big)=\text{exp}\Big(\int\displaylimits_{0}^{\alpha_s(\mu^2)}\frac{d\alpha}{\beta(\alpha,\epsilon)}\gamma^T(\alpha,n)\Big)\,.
\end{align}
Now $\gamma^T$ is the time-like DGLAP anomalous dimension. The exponentiation of anomalous dimension $\gamma^T$ reproduces all the IR divergences of the perturbative calculation, and is the $\overline{\text{MS}}$ fragmentation function, though not the physical one. Continuing the interpretation, $\Lambda$ is the location of the Landau-pole for the renormalized coupling. Finally, $\mu$ is the arbitrary factorization scale, and renormalization group evolution in $\mu$ resums the logs of $Q/\Lambda$:
\begin{align}\label{eq:DGLAP}
\mu^2\frac{d}{d\mu^2}C\Big(z,Q^2,\mu^2\Big)&=-\int_{z}^{1}\frac{dz'}{z'}\gamma^{T}\Big(\frac{z}{z'},\alpha_s(\mu^2)\Big)C\Big(z',Q^2,\mu^2\Big)\,.
\end{align}  
Through this equation one can resum the logs of $Q/\Lambda$, and $\gamma^T$ has been calculated to three loops in ref. \cite{Chen:2020uvt}, while partial results can be found in refs. \cite{Moch:2007tx,Gituliar:2015pra}. In what follows I need to resum $C,\gamma^T,\, and\,\, d_h$, not just for the logs of $Q/\Lambda$, but also logs of $\zh$ as $\zh\rightarrow 0$. To accomplish these goals, I turn to a distinct factorization of $\fxsech$ in terms of angles.

Rather than dimensional regularization, I can regulate the infra-red divergences by cutting off the angle that any other emission may approach the observed hadron, defining the angular scale $\Rir$. However I keep all quantities in $4-2\epsilon$ dimensions until after I have evolved in the angular cutoff $\Rir$. This results in a number density $\fxseca$ for the hadrons carrying $\zh$ that is a function of $\Rir$ and $Q^2$. The number density satisfies the angular-ordered evolution equation:
{\small\begin{align}
    \label{eq:evolution}
    \Rir^2\frac{d }{d\Rir^2}\zh^{1+2\epsilon} \fxseca\Big(\zh,\Rir^2,\frac{\mu^2}{Q^2}\Big)&=\rho\Big(\frac{\mu^2}{\Rir^2Q^2}\Big)\zh^{1+2\epsilon} \fxseca\Big(\zh,\Rir^2,\frac{\mu^2}{Q^2}\Big)\nonumber\\
    &\qquad+\int_{\zh}^{1}\frac{dz}{z}P\Big(\frac{\zh}{z};\frac{\mu^2}{z^2\Rir^2Q^2}\Big)z^{1+2\epsilon}\fxseca\Big(z,\Rir^2,\frac{\mu^2}{Q^2}\Big)\,.
\end{align}}
$P$ is determined by the space-like anomalous dimension in $4-2\epsilon$ dimensions, introduced in ref. \cite{Ciafaloni:2005cg,Ciafaloni:2006yk}, and $\rho$ is given by the beta function. For their detailed form, see ref. \cite{Neill:2020bwv}. Solving this equation and evolving $\Rir$ to zero resums the logs of $\zh$ in both $C$ and $\gamma^T$ of eqs. \eqref{eq:frag_factor_standard} and \eqref{eq:DGLAP}.

The fact that the number density of hadrons obeys such an evolution equation implies that I can introduce an angular-ordered fragmentation function $\ffa$ containing all emissions above the angular scale $\Rir$ up to some angular factorization scale $\Ruv$:\footnote{I call this scheme an angular factorization due to the argument of the splitting function in $4-2\epsilon$ dimensions appearing in eq. \eqref{eq:evolution}. When continuing to $4-2\epsilon$ dimensions, I continue the dimension of the transverse momentum of each loop integral. The transverse plane is defined with respect to the light-cone direction along which travels the fragmented hadron. The dimensional factor $\mu$ introduced in the regularization necessarily scales with the continued transverse momentum. Since the ratio $\frac{\mu^2}{z^2 Q^2\Rir^2 }$ appears in the splitting function of eq. \eqref{eq:evolution}, then $zQ\Rir$ must be a transverse momentum scale. Thus as $zQ$ is the energy of the parent parton to the splitting process, I interpret $\Rir$ as the angle it makes to the collinear sector defined by the direction of the fragmented particle.} 
{\small\begin{align}\label{eq:frag_factor_angle}
 \fxseca\Big(\zh,\Rir^2,\frac{\mu^2}{Q^2}\Big)&=\int_{\zh}^{1}\frac{dz}{z}  \ffa\Big(\frac{\zh}{z},\Rir^2,\Ruv^2,\frac{\mu^2}{z^2Q^2}\Big)\fxseca\Big(z,\Ruv^2,\frac{\mu^2}{Q^2}\Big)\,.
\end{align}}  
$\Ruv$ is the factorization angle. I use the notation $\Rir$ and $\Ruv$ as $\Rir$ cuts off the angle for the infra-red splittings, and $\Ruv$ cuts off the angle for the ultra-violet function. This factorization is independent of $\Ruv$, and I find that $\ffa$ again satisfies eq. \eqref{eq:evolution}. I can interpret the $\ffa$ in eq. \eqref{eq:frag_factor_angle} as simply the solution to eq. \eqref{eq:evolution} evolved between the angular scales $\Rir$ and $\Ruv$, with a delta function ($\delta(1-\zh)$) as the initial condition. Put another way, $\ffa$ is now simply the green's function for the eq. \eqref{eq:evolution}, and $\fxseca$ at $\Ruv$ is the boundary condition. The independence on $\Ruv$ follows from the semi-group property that any linear differential equation enjoys: I can evolve the initial condition straight to its end-point, or alternatively, I may stop at any intermediate time, take the solution at that time as a new initial condition, and solve again to the endpoint. The end result must be the same for any intermediate time. 

The fragmentation function $\ffa$ resulting from the factorization and evolution in terms of angles is now the fundamental object I wish to consider.

The coefficient function $C$ of \eqref{eq:frag_factor_standard} is resummed when $\Rir\rightarrow 0$ for $\ffa$ after solving eq. \eqref{eq:evolution}. Then taking $\mu=Q$,
\begin{align}\label{eq:coef_resum}
  \lim_{\Rir\rightarrow 0}\barffa\Big(n,\Rir^2,1\Big)&=\mathcal{R}^{T}(\alpha_s,n)\text{exp}\Big(\int_{0}^{\alpha_s}\frac{d\alpha}{\beta(\alpha,\epsilon)}\gamma^T(\alpha,n)\Big)\,,\\
\label{eq:coef_resum_II}  C\Big(z,\mu^2,\mu^2\Big)&=\frac{1}{z}\int\displaylimits_{c-i\infty}^{c+i\infty}\frac{dn}{2\pi i} z^{-n}\mathcal{R}^{T}(\alpha_s,n)\barfxseca\Big(n,1,1\Big)\,.
\end{align}  
Where $\mathcal{R}^{T}(\alpha_s,n)$ is simply the small-$\zh$ resummed component of the coefficient function $C\Big(z,\mu^2,\mu^2\Big)$ in moment space, and $\barfxseca\Big(n,1,1\Big)$ is the boundary condition to the evolution eq. \eqref{eq:evolution} in moment space, at $\Rir=1$ and $\mu=Q$. This boundary conditions is finite as $n\rightarrow 0$ order-by-order in fixed order perturbation theory:
\begin{align}
\lim_{n\rightarrow 0}\barfxseca\Big(n,1,1\Big)<\infty\,.
\end{align}  
As is well known, $\mathcal{R}^{T}(\alpha_s,n)$ and $\gamma^T$ are finite in the limit $n\rightarrow 0$ only after one has resummed the soft behavior, for instance, through the angular evolution equations. I introduce the notation $\mathcal{R}^{T}$ rather than just $\bar{C}$ to emphasize that this is the resummed quantity for the small-$\zh$ logarithms. Likewise, I now posit that the \emph{fragmentation function} $\ffh$ of eq. \eqref{eq:frag_factor_standard} is resummed by formally taking the limit $\Ruv\rightarrow \infty$:
\begin{align}\label{eq:fragmentation_resum}
  \lim_{\Ruv\rightarrow\infty}\barffa\Big(n,1,\Ruv^2\Big)&=\mathcal{J}^{T}(\alpha_s,n)\text{exp}\Big(-\int_{0}^{\alpha_s}\frac{d\alpha}{\beta(\alpha,\epsilon)}\gamma^T(\alpha,n)\Big)\,,\\
\label{eq:fragmentation_resum_II} \ffh(\zh,\mu^2,\mu^2)&= \frac{1}{z}\int\displaylimits_{-i\infty+c}^{i\infty+c}\frac{dn}{2\pi i} \zh^{-n}\mathcal{J}^{T}(\alpha_s,n) \,.
\end{align}  
Where $\Rir=1$, and, in an abuse of notation, I set within $\ffh$, $\Lambda=\mu$. That is to say, I dropped all terms in the fragmentation function related to the logarithmic terms determined from the anomalous dimension. These terms can be restored via the renormalization group equations. Between eqs. \eqref{eq:coef_resum} and \eqref{eq:fragmentation_resum}, the divergences associated with $\gamma^T$ will cancel. When I write eqs. \eqref{eq:coef_resum_II} and \eqref{eq:fragmentation_resum_II}, I have explicitly performed this cancellation, and defined a finite fragmentation function, suitable for a physical prediction when substituted into eq. \eqref{eq:frag_factor_standard}. When working in pure dimensional regularization calculation, the fragmentation function is the divergence of the bare calculation, which is then cast aside for a genuine model function for a comparison to the physical fragmentation spectrum.  

The treatment of the limits of the angular cut-offs are in accord with standard effective field theory techniques regarding the treatment of cutoffs: a cutoff may be introduced to define a sector of the theory, but this cutoff must be sent to either to its IR or UV limit, depending on whether the modes of the function are UV or IR, respectively.\footnote{This is a prescription that has been attempted to be formalized in the context of the method-of-regions, refs. \cite{Beneke:1997zp,Jantzen:2011nz}, a means to evaluate loop integrals in a fixed order expansion. Here I am applying it to a resummation of loop integrals.}

Of course, this relation of eq. \eqref{eq:fragmentation_resum_II} cannot be literally true for non-perturbative fragmentation, since I am working in perturbation theory and the fragmentation function is non-perturbative. However, if I were to examine rather than the fragmentation to hadrons, but semi-inclusive jet production, then such a relation would be true for the semi-inclusive jet function, which appears in the same type of the factorization theorem (eq. \eqref{eq:frag_factor_standard}) as the case of fragmentation, see refs. \cite{Kang:2016mcy,Dai:2016hzf}. They share the same coefficient function and evolution equations, differing only in that the low scale function describes how a parton forms jets via some jet algorithm, rather than hadrons.

When I claim that eq. \eqref{eq:fragmentation_resum} also resums the fragmentation function for hadrons, then I am invoking the LPHD hypothesis. This is to say that hadrons are but jets with a jet radius of the order of the mass of the hadron divided by the hard scale $Q$.

\subsection{The fragmentation function to all orders}\label{sec:ff_derive}
Going to moment space, it is straightforward to derive the resummed fragmentation function \emph{in terms of the resummed coefficient function to all orders}. To do so, I then recall the expansion of evolution kernel $P$ in eq. \eqref{eq:evolution} within dimensional regularization as worked out in ref. \cite{Neill:2020bwv}:
{\small\begin{align}\label{eq:evo_mellin}
    \Rir^2\frac{d}{d \Rir^2} \barffa(n,\Rir^2,\Ruv^2,\mu^2,Q^2)&=\sum_{\ell=1}^{\infty}\rho^{(\ell-1)}(\alpha_s;\epsilon)\Big(\frac{\mu^2}{\Rir^2Q^2}\Big)^{\ell\epsilon}\barffa\Big(n,\Rir^2,\Ruv^2,\mu^2,Q^2\Big)\nonumber\\
    &\qquad+\sum_{\ell=1}^{\infty}\bar P^{(\ell-1)}\Big(n-2\epsilon;\alpha_s;\epsilon\Big)\Big(\frac{\mu^2}{\Rir^2Q^2}\Big)^{\ell\epsilon}\barffa\Big(n-2\ell\epsilon,\Rir^2,\mu^2,Q^2\Big)\,.
\end{align}}
Setting $\mu=Q$, and writing:
{\small\begin{align}
    \barffa(n,\Rir^2,\Ruv^2,\mu^2,\mu^2)&=\barffa(n,\Rir^2,\Ruv^2)\,,
\end{align}}
then eq. \eqref{eq:evo_mellin} can be iteratively solved by introducing the functions:
{\small\begin{align}
  I\Big(\ell_1;n;\Rir,\Ruv\Big)&=\powR\int\displaylimits_{\Rir}^{\Ruv}\frac{d\theta_{1}}{\theta_1^{1+2\ell_1\epsilon}}\bar{P}^{(\ell_1-1)}(n-2\epsilon),\nonumber\\
  I\Big(\ell_1,...,\ell_k;n;\Rir,\Ruv\Big)&=\powR^k\int\displaylimits_{\Rir}^{\Ruv}\frac{d\theta_{1}}{\theta_1^{1+2\ell_1\epsilon}}\bar{P}^{(\ell_1-1)}(n-2\epsilon)\nonumber\\
  &\qquad\qquad\times\int\displaylimits_{\Rir}^{\Ruv}\prod_{i=2}^{k}\frac{d\theta_{i}}{\theta_i^{1+2\ell_i\epsilon}}\Theta(\theta_{i}-\theta_{i-1})\bar{P}^{(\ell_i-1)}\Big(n-2\epsilon(1+\sum_{j=1}^{i-1}\ell_{j})\Big)\,,\text{if } k>1\,.
\end{align}}
Then the iterative expansion is obtained to be:
{\small\begin{align}\label{eq:iterative_expansion_fin}
    \barffa\Big(n,\Rir^2,\Ruv^2\Big)&=\text{exp}\Bigg(-\int_{\Rir^2}^{\Ruv^2}\frac{d\theta^2}{\theta^2}\rho\big(\theta^{-2}\big) \Bigg)\Bigg(1+\sum_{\ell_1}I(\ell_1;n;\Rir,\Ruv)\nonumber\\
    &\qquad+\sum_{\ell_1,\ell_2}I(\ell_1,\ell_2;n;\Rir,\Ruv)+\sum_{\ell_1,\ell_2,\ell_3}I(\ell_1,\ell_2,\ell_3;n;\Rir,\Ruv)+...\Bigg)\,\,.
\end{align}}
Where all sums are from $\ell_{i}=1$ to $\infty$. Critically, since all functions are defined in dimensional regularization, I have the result:
{\small\begin{align}
  \lim_{\Rir\rightarrow\infty}\lim_{\Ruv\rightarrow 0}\barffa\Big(n,\Rir^2,\Ruv^2\Big)&=1\,.
\end{align}}
That is, all integrals are scaleless and set to zero. Then at any intermediate angle $\theta$, since $\ffa$ also satisfies eq. \eqref{eq:frag_factor_angle}, I have the result:
\begin{align}
  \lim_{\Rir\rightarrow\infty}\lim_{\Ruv\rightarrow 0}\barffa\Big(n,\Rir^2,\Ruv^2\Big)&=  \lim_{\Rir\rightarrow\infty}\lim_{\Ruv\rightarrow 0}\barffa\Big(n,\Rir^2,\theta^2\Big)\barffa\Big(n,\theta^2,\Ruv^2\Big)=1\,.\\
\lim_{\Rir\rightarrow 0}\barffa\Big(n,\Rir^2,\theta^2\Big)&=\lim_{\Ruv\rightarrow\infty}\frac{1}{\barffa\Big(n,\theta^2,\Ruv^2\Big)}\,.
\end{align}
But comparing to eqs. \eqref{eq:coef_resum} and \eqref{eq:fragmentation_resum}, setting $\theta=1$, this implies the chief result:
\begin{align}\label{eq:FF}
\mathcal{J}^{T}(\alpha_s,n)&=\frac{1}{\mathcal{R}^{T}(\alpha_s,n)}\,.
\end{align}  
The resummed fragmentation function is but the inverse in mellin space of the coefficient function. It is straightforward to check against refs. \cite{Kang:2016mcy,Dai:2016hzf} that this resummed formula indeed gives the correct logarithms for the $n_f=0$ semi-inclusive gluon jet function to the known orders.

Given this result, the coefficient function and the fragmentation function would cancel exactly to a delta function in eq. \eqref{eq:frag_factor_standard}, even after evolution, except for the fact that the running coupling is evaluated at different scales. For moderate evolution one would expect still a large cancellation, only seeing large effects when $Q^2\gg \Lambda^2$.

\subsection{Full QCD}\label{sec:full_QCD}
I discuss briefly the extension to full QCD for the angular-ordered evolution equations. The evolution equations in full QCD now become matrix equations in flavor space. I focus on the flavor singlet sector, which is inclusive and agnostic about the flavor structure of the splitting, allowing mixing between quarks and gluons. I can write for the angular factorization:
\begin{align}
\fxseca\,\!_i\Big(\zh,\Rir^2,\frac{\mu^2}{Q^2}\Big)=\sum_{j}\int_{\zh}^{1}\frac{dz}{z}  V_{ij}\Big(\frac{\zh}{z},\Rir^2,\Ruv^2,\frac{\mu^2}{z^2Q^2}\Big)\fxseca\,\!_{j}\Big(z,\Ruv^2,\frac{\mu^2}{Q^2}\Big)
\end{align}  
The index $i$ denotes either a gluon or quark singlet carrying the energy fraction $\zh$ of the scattering at angular scale $\Ruv$, the index $j$ is the flavor of the intermediate parton at angular scale $\Ruv$, and the matrix $V_{ij}$ describes the interactions when I lower the angular cut-off from $\Ruv$ to $\Rir$. For now I restrict myself to leading log accuracy in $\zh$, and I write the evolution equation for the evolution matrix in moment space as (with $\mu=Q$):
\begin{align}\label{eq:full_qcd_evo}
  \Rir^2\frac{d}{d\Rir^2}\bar{V}_{ij}(n,\Rir^2,\Ruv^2)&=\sum_{k}P_{ik}^{(0)}(n-2\epsilon)(\Rir^2)^{-\epsilon}\bar{V}_{kj}(n-2\epsilon,\Rir^2,\Ruv^2)\,.
\end{align}
where one sums over the flavors in the flavor singlet sector of the theory. Rather than the angular-ordered fragmentation function, I now have the angular-ordered fragmentation matrix. Then given an initial condition $f_{i}$, I can form a solution for the angular-ordered fragmentation function:
\begin{align}
\ffa\,\!_{i}(n,\Rir^2,\Ruv^2)&=\sum_{j}\bar{V}_{ij}(n,\Rir^2,\Ruv^2)f_{j}\,.
\end{align}  
At leading log, the splitting matrix is (in a $(g,q)$ basis for the singlet flavor space, and representing matrices in bold face):
\begin{align}
  \mathbf{P}^{(0)}(n)&=\frac{\alpha_sC_A}{\pi n}\times \mathbf{C}\,\\
  \mathbf{C}&=\Bigg (\begin{array}{cc}
   1 & \frac{C_F}{C_A} \\
   0 & 0 \\
  \end{array} \Bigg)
\end{align}
Since $\mathbf{C}^2=\mathbf{C}$, this implies that I can solve eq. \eqref{eq:full_qcd_evo} with:
\begin{align}\label{eq:full_qcd_evo_solved}
  \mathbf{\bar{V}}(n,\Rir^2,\Ruv^2)&=\barffa(n,\Rir^2,\Ruv^2)\mathbf{C}\,.
\end{align}
where $\barffa(n,\Rir^2,\Ruv^2)$ solves eq. \eqref{eq:evo_mellin} truncated to leading log with $\mu=Q$. Then the projector for the coefficient function and jet function respectively is given as:
\begin{align}\label{eq:projector}
  \mathbf{\bar{V}}_{uv}(n)&=\lim_{\Rir\rightarrow 0}\mathbf{\bar{V}}(n,\Rir^2,1)\,,\nonumber\\
  \mathbf{\bar{V}}_{ir}(n)&=\lim_{\Ruv\rightarrow \infty}\mathbf{\bar{V}}(n,1,\Ruv^2)\,. 
\end{align}

Introducing the standard factorization formula for fragmentation in full QCD:
\begin{align}\label{eq:frag_factor_standard_matrix}
  \fxsech\Big(\zh,Q^2,\Lambda^2\Big)&=\int_{\zh}^{1}\frac{dz}{z}\vec{\mathbf{C}}\Big(z,Q^2,\mu^2\Big)\cdot\vec{\mathbf{\ffh}}\Big(\frac{\zh}{z},\mu^2,\Lambda^2\Big)+...\,,
\end{align}  
then when coupled with the correct initial conditions ($f_{ir}$ and $f_{uv}$), \emph{after} dropping (canceling) the IR or UV divergences, these matrices of eq. \eqref{eq:projector} will resum the coefficient and fragmentation functions respectively:
\begin{align}
\vec{\mathbf{C}}\Big(z,\mu^2,\mu^2\Big)&=\frac{1}{z}\int\displaylimits_{c-i\infty}^{c+i\infty}\frac{dn}{2\pi i}z^{-n}\mathbf{\bar{V}}_{uv}(n)\cdot \vec{\mathbf{f}}_{uv}\,,\\
\vec{\mathbf{\ffh}}\Big(z,\mu^2,\mu^2\Big)&=\frac{1}{z}\int\displaylimits_{c-i\infty}^{c+i\infty}\frac{dn}{2\pi i}z^{-n}\vec{\mathbf{f}}_{ir}\cdot\mathbf{\bar{V}}_{ir}(n)\,.
\end{align}  
I then introduce the initial and final conditions in the $(g,q)$ basis for the resummed  $e^+e^-$ cross-section relevant for hadron production:
\begin{align}
  \vec{\mathbf{f}}_{ir}&=(1,1)\,,\\
  \vec{\mathbf{f}}_{uv}&=(0,1)\,.
\end{align}
I then have:
\begin{align}
  \fxsech(\zh,Q^2,\Lambda^2)&= \frac{1}{\zh}\int\displaylimits_{c-i\infty}^{c+i\infty}\frac{dn}{2\pi i}\zh^{-n}\vec{\mathbf{f}}_{ir}\cdot \mathbf{\bar{V}}_{ir}\Big(n,\alpha_s(\Lambda^2)\Big)\cdot\mathbf{\bar{U}}\Big(n,Q^2,\Lambda^2\Big)  \cdot\mathbf{\bar{V}}_{uv}\Big(n,\alpha_s(Q^2)\Big) \cdot \vec{\mathbf{f}}_{uv}\,\\
  &=\int_{\zh}^{1}\frac{dz}{z} F\Big(z,Q^2,\Lambda^2\Big)\frac{C_F}{C_A} \Bigg(U_{gg}\Big(\frac{\zh}{z},Q^2,\Lambda^2\Big)+\frac{C_F}{C_A} U_{qg}\Big(\frac{\zh}{z},Q^2,\Lambda^2\Big)\Bigg)\,.\\
\label{eq:frag_fun}  F\Big(z,Q^2,\Lambda^2\Big)&=\frac{1}{z}\int\displaylimits_{c-i\infty}^{c+i\infty}\frac{dn}{2\pi i}z^{-n}\frac{\mathcal{R}^{T}(\alpha_s(Q^2),n)}{\mathcal{R}^{T}(\alpha_s\big(\Lambda^2\big),n)}\,.
\end{align}  
For the order to which I am currently working, $\mathcal{R}^{T}$ will given by eq. \eqref{eq:ll_resum_coef}. $\mathbf{\bar{U}}\Big(n,Q^2,\Lambda^2\Big)$ is the matrix solution of eq. \eqref{eq:DGLAP} when generalized to multiple flavors in moment space, using the small-$\zh$ resummed time-like splitting functions that can be found for the full singlet case in ref. \cite{Kom:2012hd}.

I have written the hard projected state as:
\begin{align}\label{eq:projected}
\mathbf{\bar{V}}_{uv}(n)\cdot\vec{\mathbf{f}}_{uv}=\Big(\frac{C_F}{C_A}\mathcal{R}^{T}(\alpha_s,n),0\Big)\,.
\end{align}  
One may wonder why the initial state is not:
\begin{align}\label{eq:curiou}
\mathbf{\bar{V}}_{uv}(n)\cdot\vec{\mathbf{f}}_{uv}=\Bigg(\frac{C_F}{C_A}\Big(\mathcal{R}^{T}(\alpha_s,n)-1\Big),0\Bigg)\,.
\end{align}  
Strictly speaking, if I expand eq. \eqref{eq:curiou}, I reproduce perturbation theory. If I expand eq. \eqref{eq:projected}, I will get a term proportional to $\delta(1-z)$ that does not appear in the explicit calculation of the $e^+e^-$ coefficient function. The answer lies in the numerical evaluation of the contour integral performing the mellin inversion, in say eq. \eqref{eq:frag_fun}. As long as I stay strictly away from $z=1$ exactly, I never pick up the delta function that appears \emph{if} I were to perform the perturbative expansion and then mellin invert. Thus I automatically implement the delta function subtraction that eq. \eqref{eq:curiou} explicitly implements.

\section{NLL versus MLLA}
Going back to pure Yang-Mills, I am now in a position to write down the resummed spectrum to all orders. Once I derive the resummed anomalous dimension $\gamma^T$ and coefficient function $\mathcal{R}^T$, I have:
{\small\begin{align}\label{eq:resum}
  \fxsech\Big(\zh,Q^2,\Lambda^2\Big)&=\frac{1}{\zh}\int\displaylimits_{c-i\infty}^{c+i\infty}\frac{dn}{2\pi i}\zh^{-n}\frac{\mathcal{R}^{T}(\alpha_s(Q^2),n)}{\mathcal{R}^{T}(\alpha_s(\Lambda^2),n)}\barfxseca\Big(n,1,1\Big)\text{exp}\Bigg(\int\displaylimits^{Q^2}_{\Lambda^2}\frac{d\mu^2}{\mu^2}\gamma^T\Big(n,\alpha(\mu^2)\Big)\Bigg)
\end{align}  }
All singular $n$-dependence has been resummed into the functions $\gamma^T$ and $\mathcal{R}^T$, and the factor $\barfxseca\Big(n,1,1\Big)$ is the UV boundary condition to the angular evolution, to be evaluated at the scale $\mu=Q$, and cannot be deduced from the evolution eq. \eqref{eq:evolution} or the function $\ffa$ in eq. \eqref{eq:frag_factor_angle}. It must be obtained from a matching calculation. It formally holds in pure-Yang Mills theory to N$^2$LL accuracy that $\barfxseca\Big(n,1,1\Big)=1$, counting $\alpha_s\sim n^2$, see ref. \cite{Neill:2020bwv}.

To compare to the MLLA, I content myself to LL accuracy in the coefficient and fragmentation function. This results in an effective NLL accuracy for the resummed cross-section, so that I call eq. \eqref{eq:resum} NLL accurate when $\mathcal{R}^T$ is truncated to leading (in small $\zh$ log counting) accuracy due to the ratio-form the resummed coefficient and jet functions take. The ratio form implies a cancellation at fixed order, such that the expanded ratio is down by a power of $\alpha_s$ than it would naively be when $\alpha_s$ is taken at a common scale. Thus when the anomalous dimension is taken at NLL level for the small-$\zh$ logs, and the coefficient and fragmentation functions are taken at LL in the small-$\zh$ counting, I call the whole formula NLL. Of course, I could take all resummed quantities to be NLL in the small-$\zh$ counting, and this is perhaps the more consistent thing to do, but I leave such investigations to later work.

The MLLA result is written as:
\begin{align}
\fxsech\Big(\zh,Q^2,\Lambda^2\Big)&=\frac{1}{\zh}\int\displaylimits_{c-i\infty}^{c+i\infty}\frac{dn}{2\pi i}\zh^{-n}B(n,\Lambda^2)\text{exp}\Bigg(\int\displaylimits^{Q^2}_{\Lambda^2}\frac{d\mu^2}{\mu^2}\gamma^{MLLA}\Big(n,\alpha(\mu^2)\Big)\Bigg)
\end{align}  
Where $\gamma^{MLLA}$ is the mixed leading log anomalous dimension, given as (refs. \cite{Fong:1990nt,Dokshitzer:1991ej}):
\begin{align}
\gamma_0\Big(\frac{a_s}{n^2}\Big)&=\frac{1}{2}\Big(-1+\sqrt{1+8\frac{\alpha_sC_A}{\pi n^2}}\Big)\,,\label{eq:ll_anom_dim}\\
\gamma^{MLLA}&=\frac{1}{2}n\gamma_0-\frac{11}{24}n^2\gamma_0^2\Big(\frac{(1+\gamma_0)^3}{(1+2\gamma_0)^2}\Big)\,.
\end{align}  
However, the DGLAP splitting kernel to NLL is:
\begin{align}
   \gamma^T&=\frac{1}{2}n\gamma_0+\frac{11}{24}n^2\gamma_0^2\Big(\frac{1+\gamma_0}{1+2\gamma_0}\Big)-\frac{11}{12}\frac{\alpha_s C_A}{\pi}+...\,.
\end{align}  
Clearly, these are in-equivalent. However, the MLLA formula assumes that all logs are explicitly exponentiated via the anomalous dimension, while the NLL cross-section has implicit logs of $Q^2$ and $\Lambda^2$ in the fragmentation and coefficient functions via the running of $\alpha_s$. This suggests that to compare to MLLA, I ought to also force the exponentiation of the fragmentation and coefficient functions, so I write using the leading log resummation of $\mathcal{R}^{T}$:
\begin{align}\label{eq:ll_resum_coef}
  \mathcal{R}^{T}&=(1+2\gamma_0)^{-1/2}\,,\\
\text{ln}\frac{\mathcal{R}^{T}(\alpha_s(Q^2),n)}{\mathcal{R}^{T}(\alpha_s(\Lambda^2),n)}&\approx \frac{11}{24}n^2\gamma_0^2\Big(\frac{(1+\gamma_0)^2}{(1+2\gamma_0)^2}\Big)\text{ln}\frac{Q^2}{\Lambda^2}+...\,.
\end{align}  
I have used the renormalization group equation for $\alpha_s$ to rewrite:
\begin{align}
\alpha_s(\Lambda^2)&=\alpha_s(Q^2)+\frac{11}{12} \frac{\alpha_s^2(Q^2)C_A}{\pi}\text{ln}\frac{Q^2}{\Lambda^2}+...
\end{align}
Note that I truncated the solution to the RG equation to the first non-trivial order in $\alpha_s$, and made a similar approximation in $\mathcal{R}^{T}(\alpha_s(\Lambda^2),n)$ once I redefined the running coupling. This is an unreliable approximation to the leading log running when $\Lambda^2\ll Q^2$, never mind subleading logs. Pressing on none-the-less, I have:
\begin{align}
\gamma^{MLLA}-\Big(\gamma^T+Q^2\frac{d}{dQ^2}\text{ln}\frac{\mathcal{R}^{T}(\alpha_s(Q^2),n)}{\mathcal{R}^{T}(\alpha_s(\Lambda^2),n)}\Big)\approx -\frac{11}{12}\frac{\alpha_s C_A}{\pi}\Bigg(1+O\Big(\alpha_s\text{ln}\frac{Q^2}{\Lambda^2}\Big)\Bigg)\,,
\end{align}  
where in the formula, I assume that $\alpha_s$ is evaluated at the scale $Q^2$. Note that the difference with the MLLA is entirely due to running coupling effects being truncated in an unjustified fashion. Since the difference is independent of $n$ to the order I truncated, the difference will be just an adjustment to the scale evolution of the normalization of the spectrum, \emph{if I assume I can truncate the running coupling as I do}. That is, the collinear log $\text{ln}\frac{Q^2}{\Lambda^2}$ and its small-$\zh$ resummed coefficient, is handled correctly only for moderate evolution, and beyond moderate evolution, one expects a departure from MLLA.

\section{Parton-Hadron Duality}
In this section I extend the small-$\zh$ resummed fragmentation function into a genuine model for charged hadron production. First I illustrate the basic features of the model, built on the assumptions that the coupling freezes in the IR, and that the shape of the fragmentation function is given by the small $\zh$-function from resummed perturbation theory. Moreover, I assume that the fragmentation spectrum is dominated by its behavior at small $\zh$, so that naively forcing momentum conservation (a power correction) coupled with iso-spin symmetry dictates the evolution of the normalization. This enforcing of the normalization could be dispensed with if I matched the small-$\zh$ resummation to the full DGLAP equations, which would then have momentum conservation built in from the beginning, but I leave that for later work.

\subsection{Fragmentation model from ln$\zh$ resummation}
Here I propose that according to the LPHD hypothesis, the shape of the fragmentation function \emph{is} simply the resummed jet function with an effective $\alpha_s$ coupling in the infra-red:
\begin{align}
\barffh\Big(n,\mu^2,\Lambda^2\Big)&=\frac{1}{\mathcal{R}^{T}(\alpha^{eff}(\mu^2,\Lambda^2),n)}\Big(1+\text{anomalous dimension terms}\Big)\,.
\end{align}  
I have omitted the terms generated by the anomalous dimension that vanish when $\mu^2=\Lambda^2$. For my purposes, I adopt the simple effective coupling by shifting the argument of $\alpha_s$ by a constant mass scale, thus freezing the running coupling at a quasi-perturbative scale:
\begin{align}
\alpha^{eff}(\mu^2,\Lambda^2)=\alpha_s(\mu^2+c\Lambda^2)\,.  
\end{align}
I have also checked that using a hard cutoff in the running coupling, stopping it at a fixed mass scale, also works as a non-perturbative model, yielding decent fits to the data. In total, this implies I have 2 degrees of freedom in the model for the fragmentation function: the scale I freeze the coupling given by $c$,  and finally the scale I stop DGLAP evolution, the scale of confinement $\Lambda^2$. While this effective coupling is strictly a guess that at the form of the non-perturbative dynamics, in the spirit of LPHD, \emph{the form of the function is completely determined by the small-$\zh$ resummation, and was calculated in perturbation theory}. There will be limitations to such a fragmentation function, particularly in the extremely small-$\zh$ limit, $\zh\sim \frac{m_h^2}{Q^2}$, when the mass of the hadron $m_h$ causes significant changes to the dispersion relation between momentum and energy. Further it is expected to be wrong in the threshold region $\zh\rightarrow 1$.

In pure Yang-Mills, I write the prediction for the fragmentation spectrum as:
{\small\begin{align}\label{eq:frag_resum}
  \fxsech\Big(\zh,Q^2,\Lambda^2\Big)&=\frac{1}{\zh N(Q^2,\Lambda^2)}\int\displaylimits_{-i\infty+c}^{i\infty+c}\frac{dn}{2\pi i}\zh^{-n}\frac{\mathcal{R}^{T}(\alpha_s(Q^2+c\Lambda^2),n)}{\mathcal{R}^{T}(\alpha_s\big((1+c)\Lambda^2\big),n)}\text{exp}\Bigg(\int\displaylimits^{Q^2}_{\Lambda^2}\frac{d\mu^2}{\mu^2}\gamma^T\Big(n,\alpha(\mu^2+c\Lambda^2)\Big)\Bigg)\,.
\end{align}  }
Note that for consistency sake I shift the argument of the coupling everywhere, but in practice this does not matter for the high-scale function or anomalous dimension. Further I have suppressed the initial condition for the angular evolution, taking it to be 1 in moment space, which is valid to N$^2$LL accuracy. $N(Q^2,\Lambda^2)$ is a function dictated by the demand the fragmentation spectrum should obey the sum-rule:
 \begin{align}\label{eq:sum_rule}
1=\int_{0}^{1} dz z \fxsech\Big(z,Q^2,\Lambda^2\Big)\,.
\end{align}
This holds since $\fxsech$ is already the fragmentation spectrum to any hadron. This implies that the normalizing factor is:
\begin{align}
N(Q^2,\Lambda^2)=\frac{\mathcal{R}^{T}(\alpha_s(Q^2+c\Lambda^2),1)}{\mathcal{R}^{T}(\alpha_s\big((1+c)\Lambda^2\big),1)}\text{exp}\Bigg(\int\displaylimits^{Q^2}_{\Lambda^2}\frac{d\mu^2}{\mu^2}\gamma^T\Big(1,\alpha_s(\mu^2+c\Lambda^2)\Big)\Bigg)\,.
\end{align}  
As far as the expansion in $\zh$ is concerned, this is a perfectly valid modification to the spectrum, since formally these are power corrections that go beyond the small-$\zh$ approximation, and does not therefore invalidate the logarithmic counting or changes the shape of the spectrum.

Finally since I will be comparing to fragmentation to charged hadron data, I must introduce a normalization factor $N_{h^\pm}$, since the normalization can no longer be fixed by the sum-rule itself:
{\small\begin{align}\label{eq:frag_resum_II}
  \fxsech_{\pm}\Big(\zh,Q^2,\Lambda^2\Big)&=\frac{N_{h^\pm}}{\zh N(Q^2,\Lambda^2)}\int\displaylimits_{-i\infty+c}^{i\infty+c}\frac{dn}{2\pi i}\zh^{-n}\frac{\mathcal{R}^{T}(\alpha_s(Q^2+c\Lambda^2),n)}{\mathcal{R}^{T}(\alpha_s\big((1+c)\Lambda^2\big),n)}\text{exp}\Bigg(\int\displaylimits^{Q^2}_{\Lambda^2}\frac{d\mu^2}{\mu^2}\gamma^T\Big(n,\alpha(\mu^2+c\Lambda^2)\Big)\Bigg)\,.
\end{align}  }

Enforcing the sum while still restricted to charged hadron data is a phenomenological motivated modification that is based on the following assumptions:
\begin{itemize}
\item That I will predict the \emph{shape} of the spectrum correctly for all Q.
\item  Momentum is conserved, and the shape of the spectrum of neutral hadrons is identical to charged hadrons. In full QCD, this assumption hinges on the extended iso-spin symmetry governing up, down and strange quarks, since light hadron production is dominated by nucleons, pions, and kaons, and thus can be viewed as combining the LPHD hypothesis with iso-spin symmetry.
\item The spectrum is suppressed at $\zh\rightarrow 1$, and the small $\zh$ region starts at $\zh\sim 0.5$, so the fact that the fragmentation function is wrong above this point is of little consequence, as long as it is small. Thus enforcing the sum-rule on the small-$\zh$ resummed cross-section gives the correct evolution of the normalization, as this is the region where the bulk of the momentum resides.  
\end{itemize}

\section{Numerics}
\subsection{Implementation}
I evolve the DGLAP equations in momentum space, using a Runge-Kutta based discretization scheme detailed in app. \ref{sec:numerics_notes}. I thus avoid any diagonalization of matrices as must be performed for mellin space evolution. The appropriate spectrum for $e^+e^-$ is calculated as:
\begin{align}
    \fxsech_{\pm}\Big(\zh,Q^2,\Lambda^2\Big)&=\frac{N_{h^{\pm}}}{N(Q^2,\Lambda^2)}\int_{\zh}^{1}\frac{dz}{z} F\Big(z,Q^2,\Lambda^2\Big)\Bigg(U_{gg}\Big(\frac{\zh}{z},Q^2,\Lambda^2\Big)+\frac{C_F}{C_A} U_{qg}\Big(\frac{\zh}{z},Q^2,\Lambda^2\Big)\Bigg)\,.
\end{align}  
The function $F$ is given by eq. \eqref{eq:frag_fun}, and $U_{ij}$ is calculated by starting the Runge-Kutta evolution (in the notation of app. \ref{sec:numerics_notes}) with the initial conditions $H_{g}(z,t=0)=\delta(1-z), H_q(z,t=0)=0$, and evolving with the full NLL small-$\zh$ resummed singlet splitting matrix given in ref. \cite{Kom:2012hd}. At the b-quark threshold I switch to a 4 flavor scheme for the splitting functions and $\alpha_s$, though remaining in a 5 flavor scheme makes little difference in the fits. To compute $F$, I deform the integration contour into a staple as follows: first I define $n_0=\epsilon+\sqrt{\frac{\alpha_s((1+c)\Lambda^2)C_A}{\pi}}$, where $\epsilon$ is a small number. Then $\pm in_0$ is approximately where the cut furthest from the origin begins in the integrand. The cuts corresponding to $n\sim \sqrt{\alpha_s}$ with $\alpha_s$ at hard scale are closer to the origin and bounded by the staple automatically. I deform the mellin inversion contour into a staple shape surrounding the cuts, stretching from $-\infty-in_0$, to $\epsilon-in_0$, then up to $\epsilon+in_0$, and from there back to $-\infty+in_0$. All integrals on each leg of the staple are then nicely numerically convergent. I then numerically compute the convolution, and derive $N(Q^2,\Lambda^2)$ by enforcing the sum rule in eq. \eqref{eq:sum_rule}. Finally, I use the two-loop running of $\alpha_s$, with $\alpha_s=0.1187$ at the Z-pole.

\begin{figure}\center
   \hspace{-10pt} \includegraphics[width=0.45\textwidth]{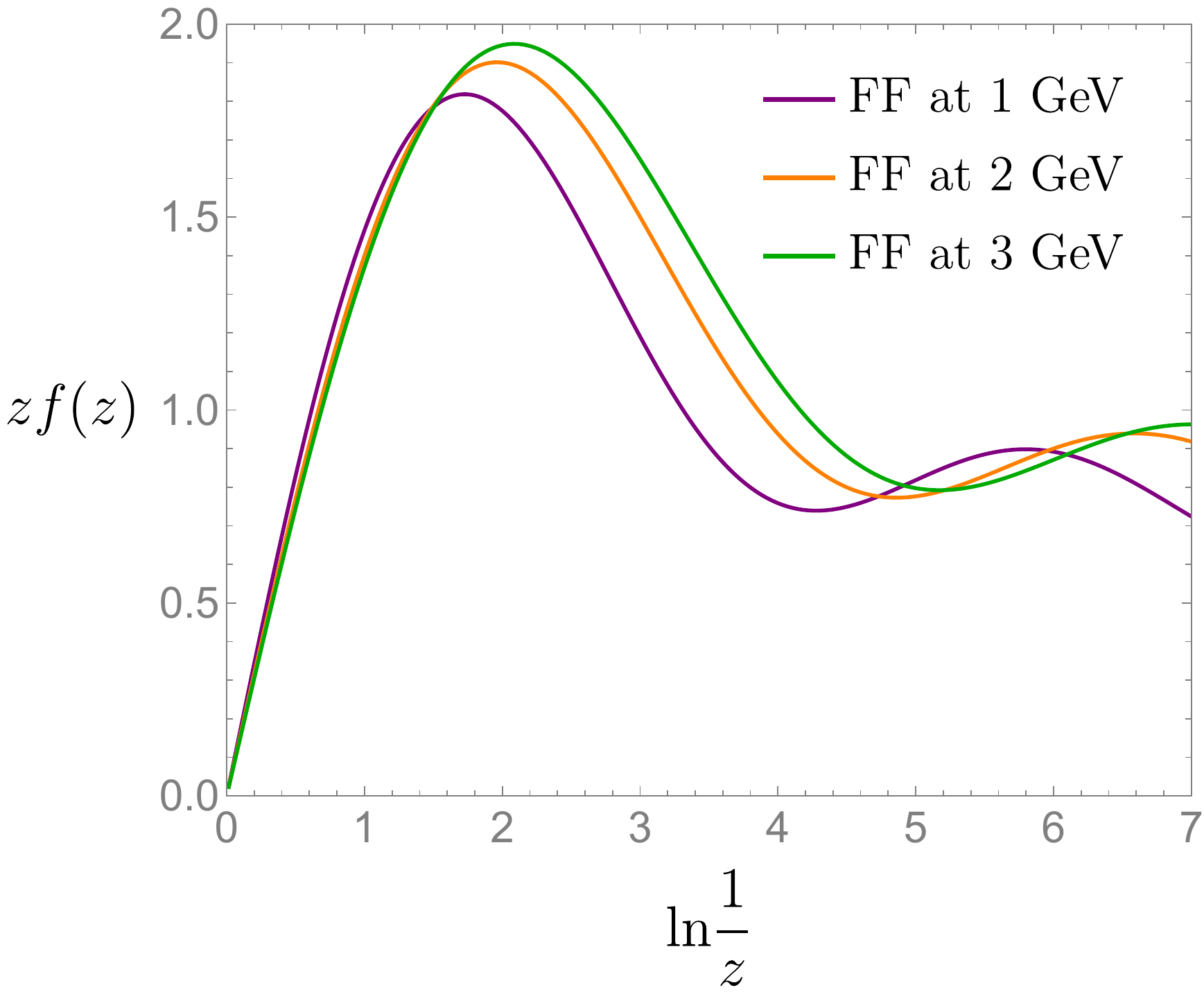}    \includegraphics[width=0.45\textwidth]{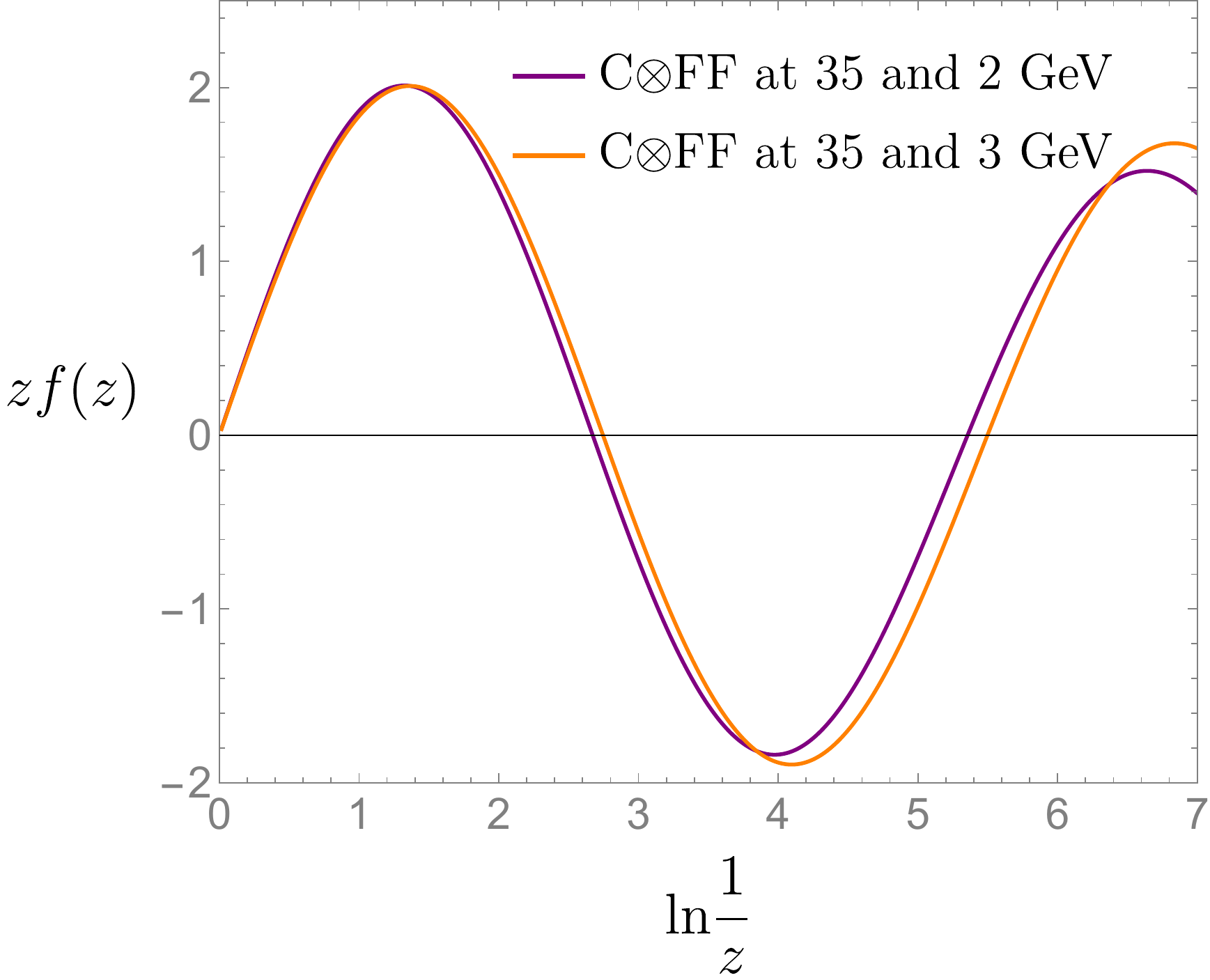}
   \caption{\label{fig:FF} On the left, I give the resummed fragmentation function of eq. \eqref{eq:FF} in momentum space evaluated at $\alpha_s(1 \text{GeV})$, $\alpha_s(2 \text{GeV})$, and $\alpha_s(3 \text{GeV})$. On the right, I give the convolution of the resummed fragmentation function with the resummed coefficient function, defined in eq. \eqref{eq:frag_fun}, with a hard scale of $35$ GeV, and fragmentation scales of $2$ and $3$ GeV.}
 \end{figure}

To gain some intuition about the predicted shape of the fragmentation function I derived in this paper, in fig. \ref{fig:FF} I plot the numerical form of the fragmentation function derived in sec. \ref{sec:ff_derive} where the resummation is performed to leading log accuracy for small energy fraction logs. Thus $\mathcal{R}_T$ is defined with eqs. \eqref{eq:ll_resum_coef} and \eqref{eq:ll_anom_dim}. I also present the results for the convolution of the fragmentation function with the coefficient function at a hard reference scale of $35$ GeV. Both plots are normalized according to the sum rule in eq. \eqref{eq:sum_rule}.

\subsection{Results}

 \begin{figure}\center
   \hspace{-10pt} \includegraphics[width=0.45\textwidth]{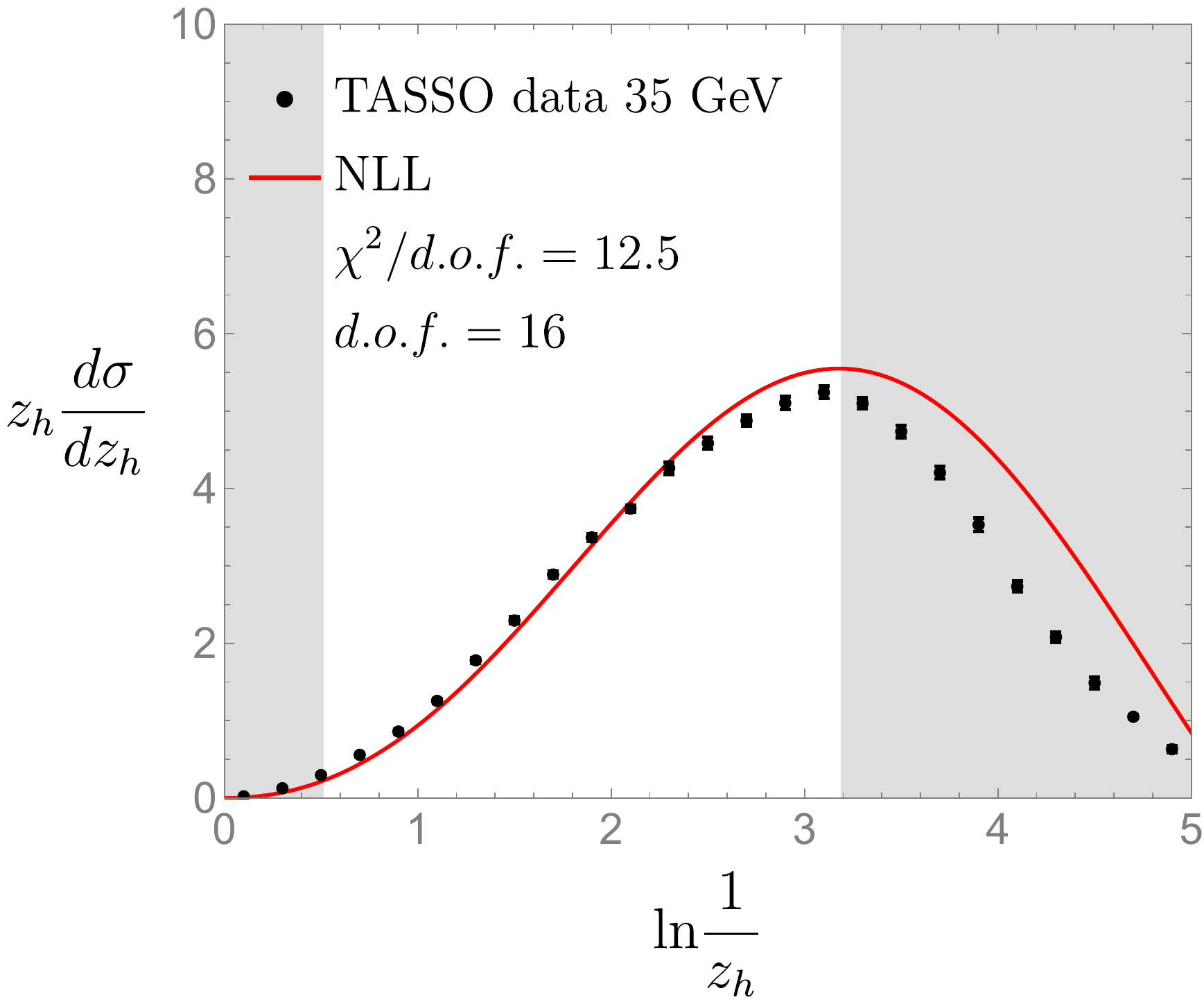}   \includegraphics[width=0.45\textwidth]{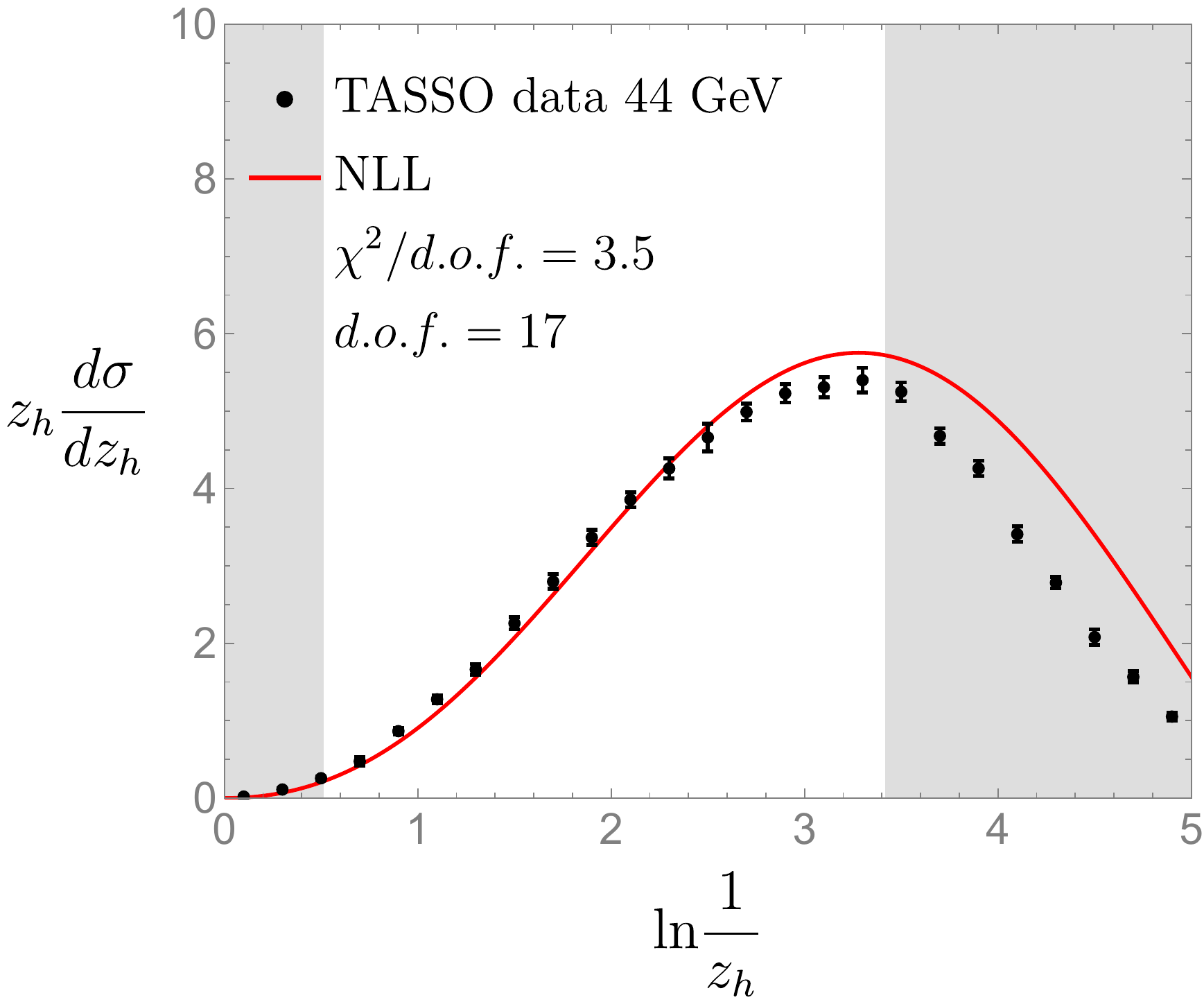} 
   \caption{\label{fig:low_Q} The fragmentation spectrum for $e^+e^-\rightarrow h^{\pm}+X$ at center of mass energies $Q=35$ and $44$ GeV. The gray region is excluded from the \chidof determination based on hadron mass corrections or $\zh\sim O(1)$, and the quoted \chidof is the goodness-of-fit for that data set alone using the best fit parameters.}
 \end{figure}

 \begin{figure}\center

  \includegraphics[width=0.45\textwidth]{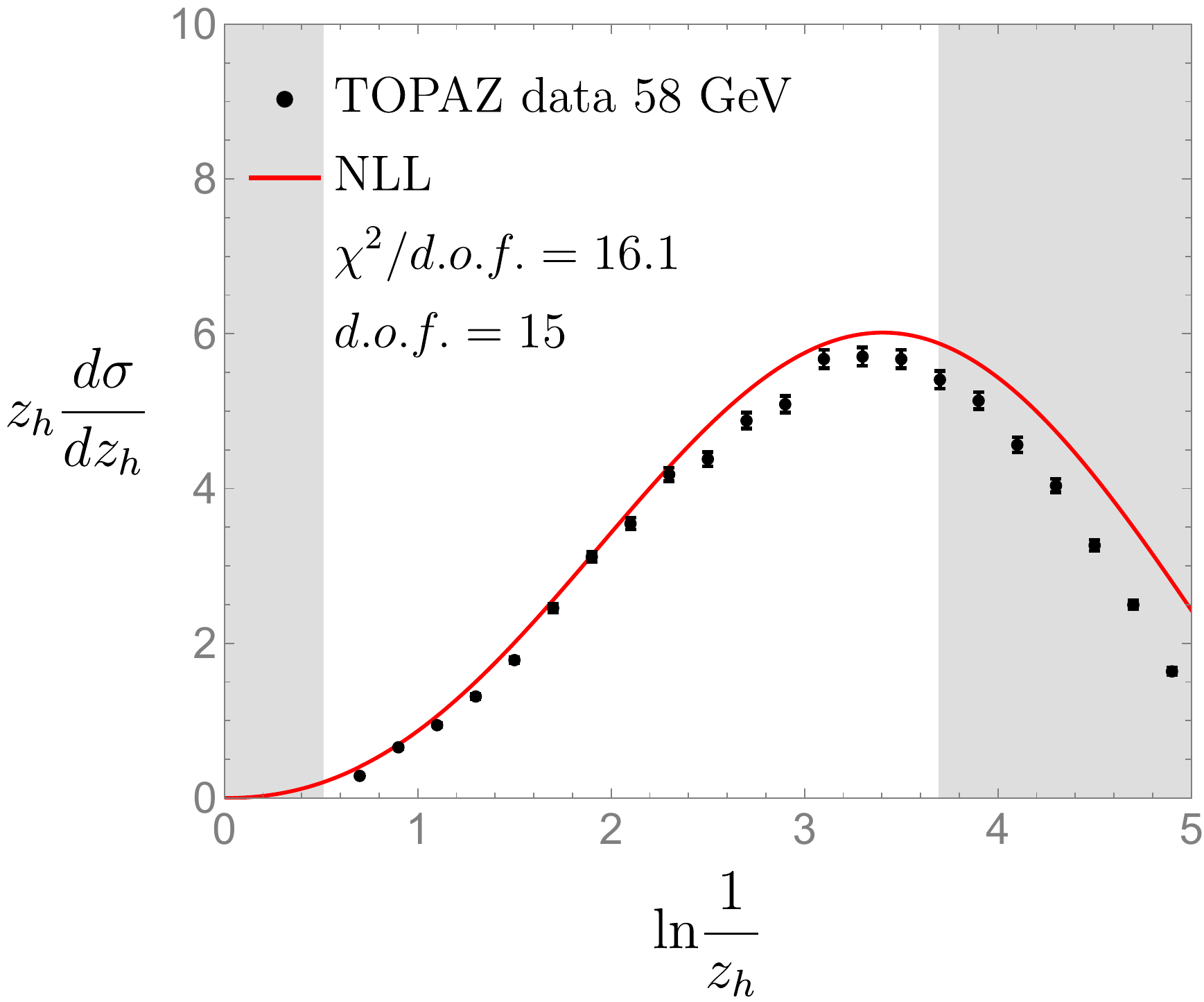}  \includegraphics[width=0.45\textwidth]{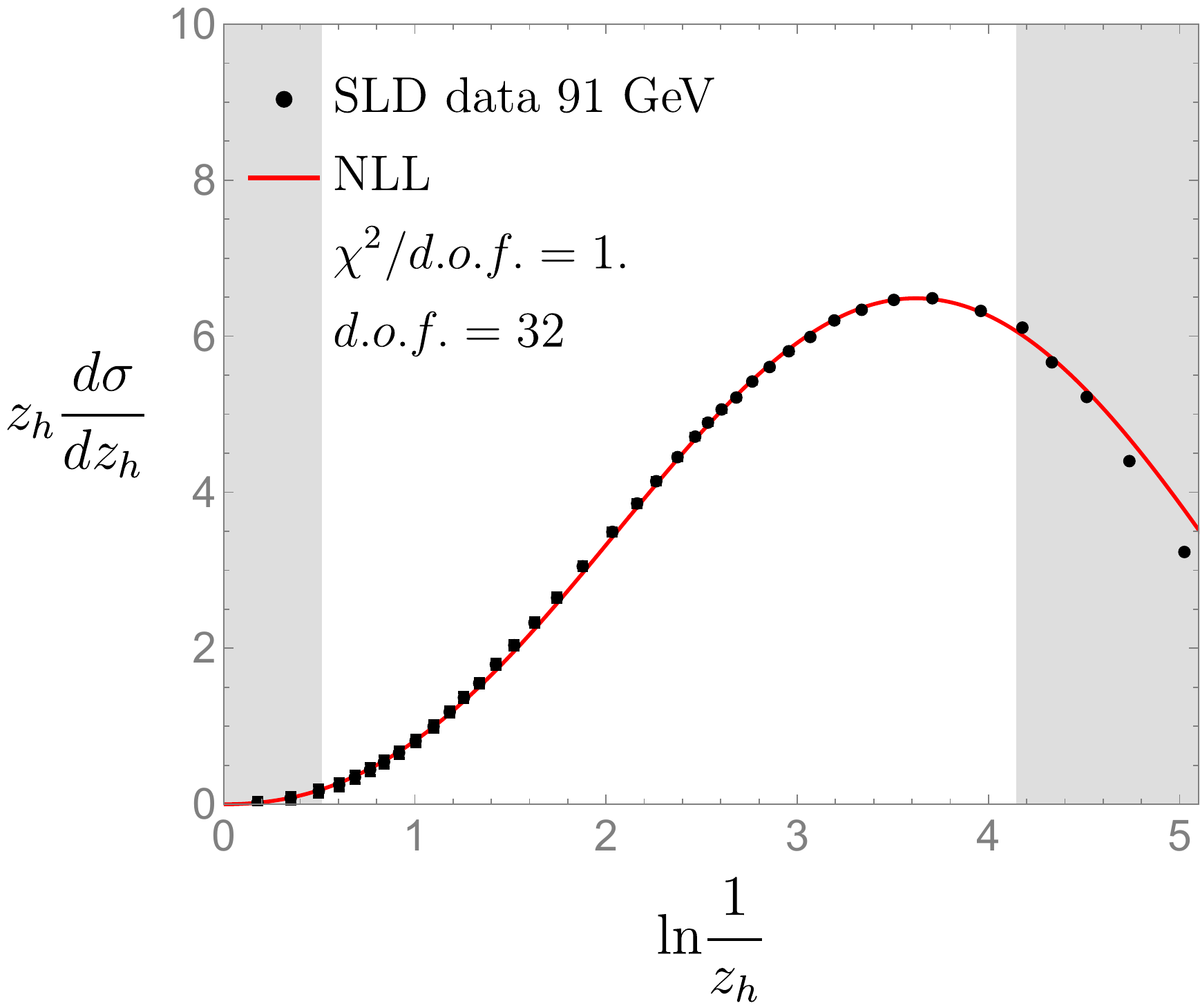} 
   \caption{\label{fig:mid_Q} The fragmentation spectrum for $e^+e^-\rightarrow h^{\pm}+X$ at center of mass energies $58$ and $91$ GeV. The gray region is excluded from the \chidof determination based on hadron mass corrections or $\zh\sim O(1)$, and the quoted \chidof is the goodness-of-fit for that data set alone using the best fit parameters.}
 \end{figure}  

 \begin{figure}\center
   \includegraphics[width=0.45\textwidth]{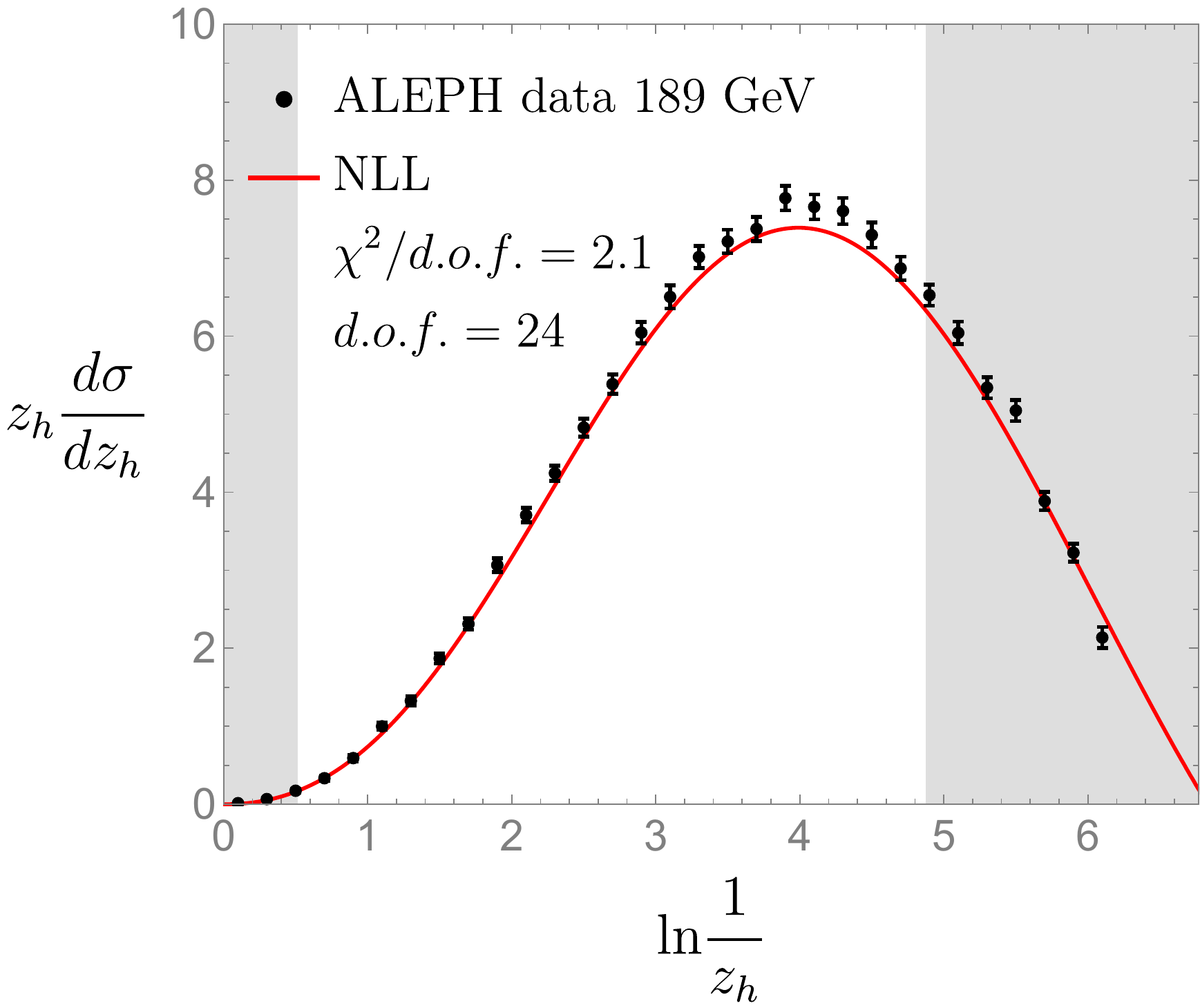}   \includegraphics[width=0.45\textwidth]{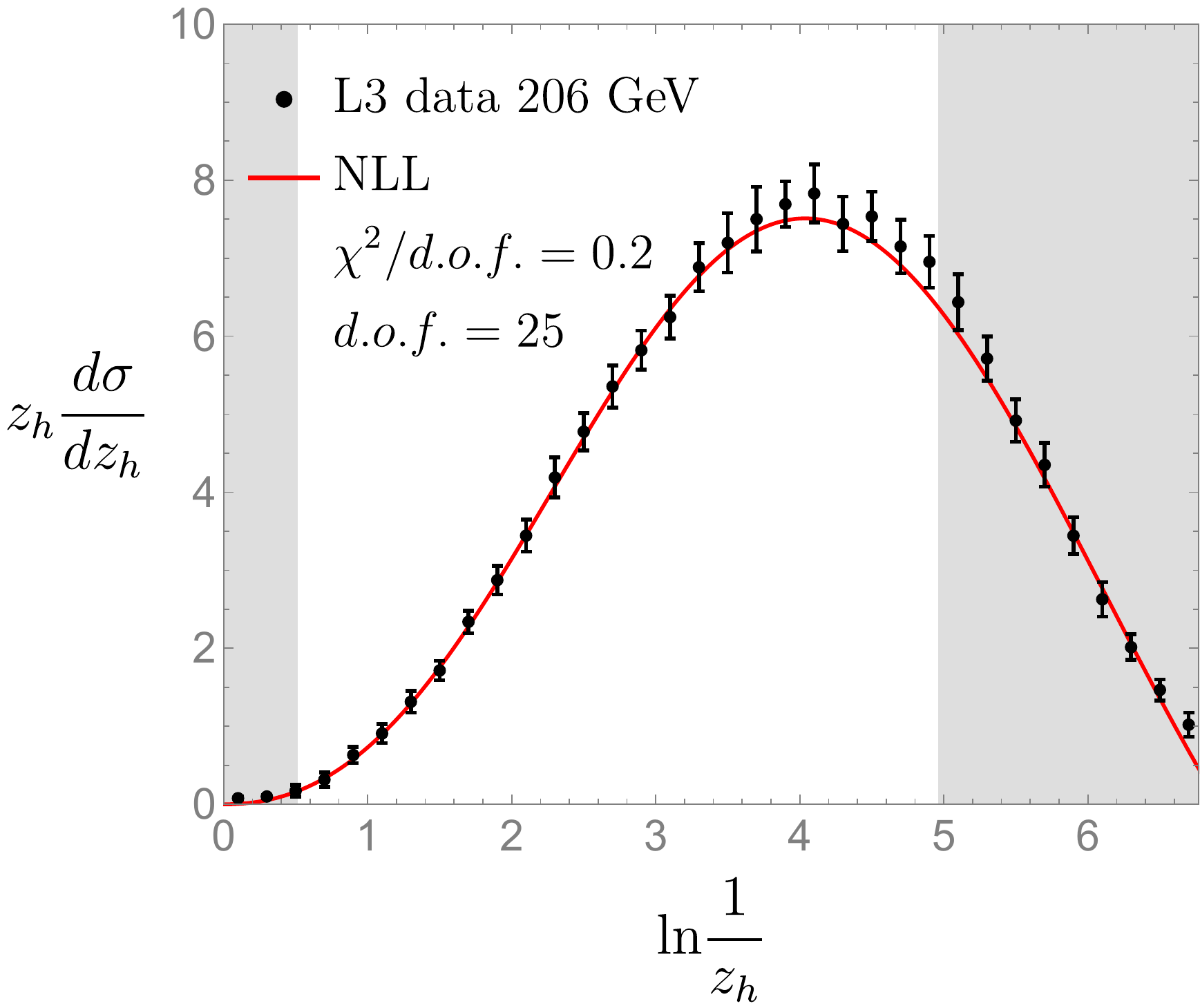}
   \caption{\label{fig:high_Q} The fragmentation spectrum for $e^+e^-\rightarrow h^{\pm}+X$ at center of mass energies $189$ and $206$ GeV. The gray region is excluded from the \chidof determination based on hadron mass corrections or $\zh\sim O(1)$, and the quoted \chidof is the goodness-of-fit for that data set alone using the global best fit parameters.}
 \end{figure}  

 I compare the single inclusive annihilation data to charged hadrons, $e^+e^-\rightarrow h^{\pm}+X$, for the following experiments:
\begin{itemize}
\item TASSSO data at $Q=35\text{GeV} , 44$GeV \cite{Braunschweig:1990yd}, in regions $0.5<-\text{ln}\zh<3.19$, $0.5<-\text{ln}\zh<3.42$, respectively, in fig. \ref{fig:low_Q}.
\item TOPAZ data at $Q=58$GeV \cite{Itoh:1994kb}, in region $0.5<-\text{ln}\zh<3.69$,  in fig. \ref{fig:mid_Q}.
\item SLD data at the Z-pole of $Q=91$GeV \cite{Abe:2003iy}, in region $0.5<-\text{ln}\zh<4.14$, in fig. \ref{fig:mid_Q}.
\item ALEPH data at $Q=189$GeV \cite{Heister:2003aj}, in region $0.5<-\text{ln}\zh<4.87$, in fig. \ref{fig:high_Q}.
\item L3 data at $Q=206$GeV \cite{Achard:2004sv}, in region $0.5<-\text{ln}\zh<4.96$, in fig. \ref{fig:high_Q}.
\end{itemize}
The upper end of these regions is determined by when I expect hadron mass corrections to become significant. To estimate this, I compare the energy fraction ($\zh$) versus the 3-momentum fraction of the hadron ($x_p$), whose relation is given by: $x_p=\zh-\frac{2m_h^2}{\zh Q^2}+O(Q^4)$, with $m_h$ the hadron mass (ref. \cite{Nason:1993xx}). Since I am fitting for generic light hadron production, I take $m_h\sim 0.5$ GeV, the mass of the kaon, and demand that $|x_p-\zh|<0.2$. More stringent criteria tends to improve the \chidof for the fit. For comparison to the limiting spectrum of the MLLA approximation, I also compare to OPAL data at $Q=91$GeV, which was fitted to the limiting spectrum, and which is largely consistent with the SLD data, though with larger error. This data I do not include in any of the $\chi^2$ determinations. The total $\chi^2$ per degree of freedom (\chidof) is 4.6, however, the fit is completely dominated by the SLD data due to the extremely small error bars quoted. Restricted to SLD, ALEPH, and L3 data alone the  \chidof is 1.1. Finally, I note that the SLD data is consistent with the earlier OPAL data of ref. \cite{Akrawy:1990ha}, and ALEPH data at $Q=206$ GeV is consistent with the L3 data at the same energy.

 Though the above quoted regions contain the points I use for the $\chi^2$ determination, I note that qualitatively the fits extrapolate essentially for all $\zh$, and this is even true \emph{quantitatively} for the data at the Z-pole and above. For instance, if I compute the \chidof on the L3 data at $206$ GeV alone, I obtain 0.3 for all data points between $0.5<-\text{ln}\zh<6.6$ and similarly for the ALEPH data where I would obtain a \chidof of 2.5 in the same range, without changing the fit.

 I find that the best fit values are obtained with $N_{h^{\pm}}=1.188, c=1.0, \Lambda = 1.0$ GeV. I note that the TASSO data at lower $Q^2$ data prefers a larger value of $\Lambda$, at $2$ GeV, but the same normalization. I speculate as to reasons for this in sec. \ref{sec:lowq}. I note that I can obtain good fits to the data with $c=2.0$ or $c=0.5$ if I also change the end point of the splitting evolution, and one could also explore attempting to fit the value of $\alpha_s$ as well.

\subsection{Comparison to MLLA} 
\begin{figure}\center
   \hspace{-10pt} \includegraphics[width=0.45\textwidth]{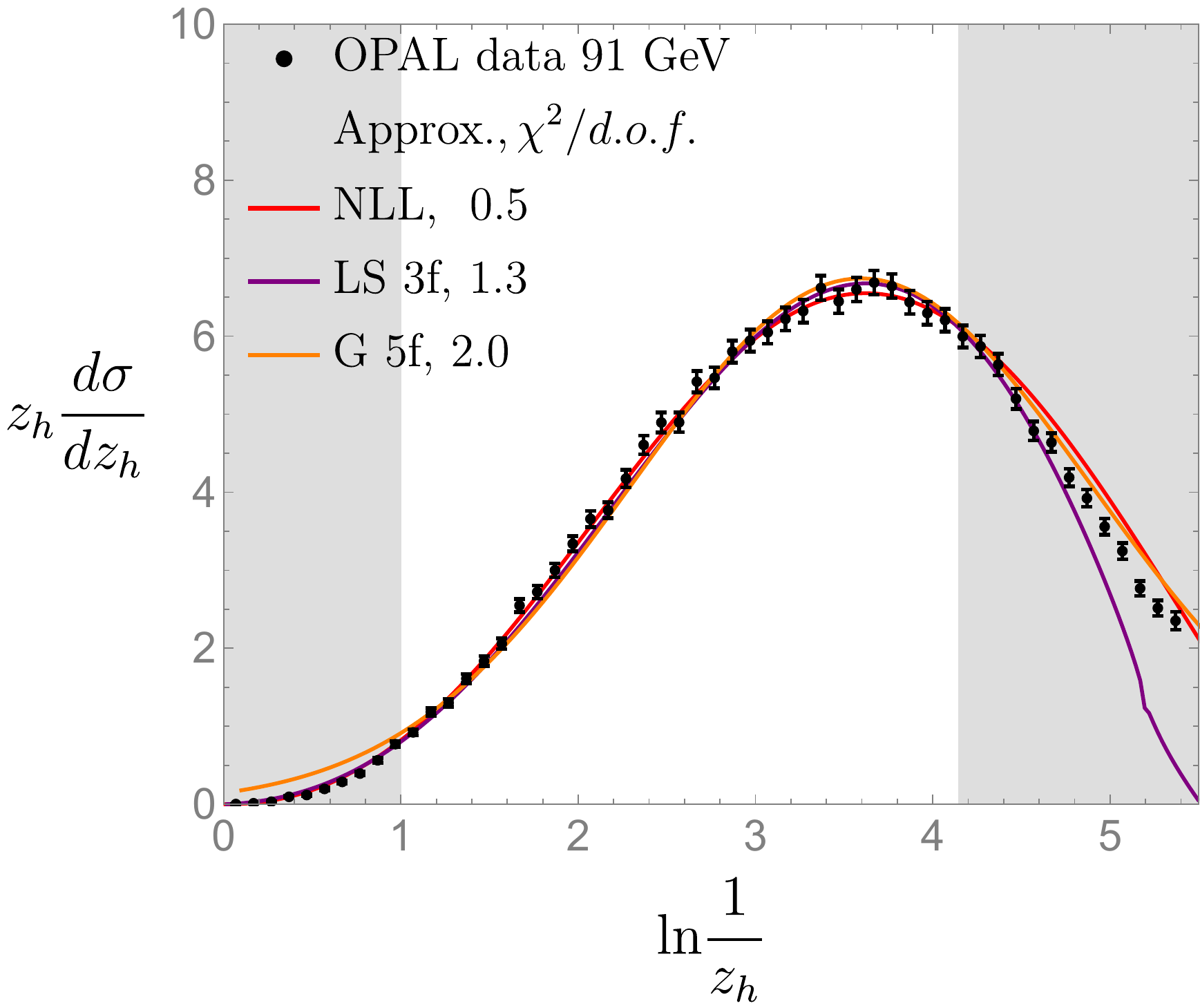}    \includegraphics[width=0.45\textwidth]{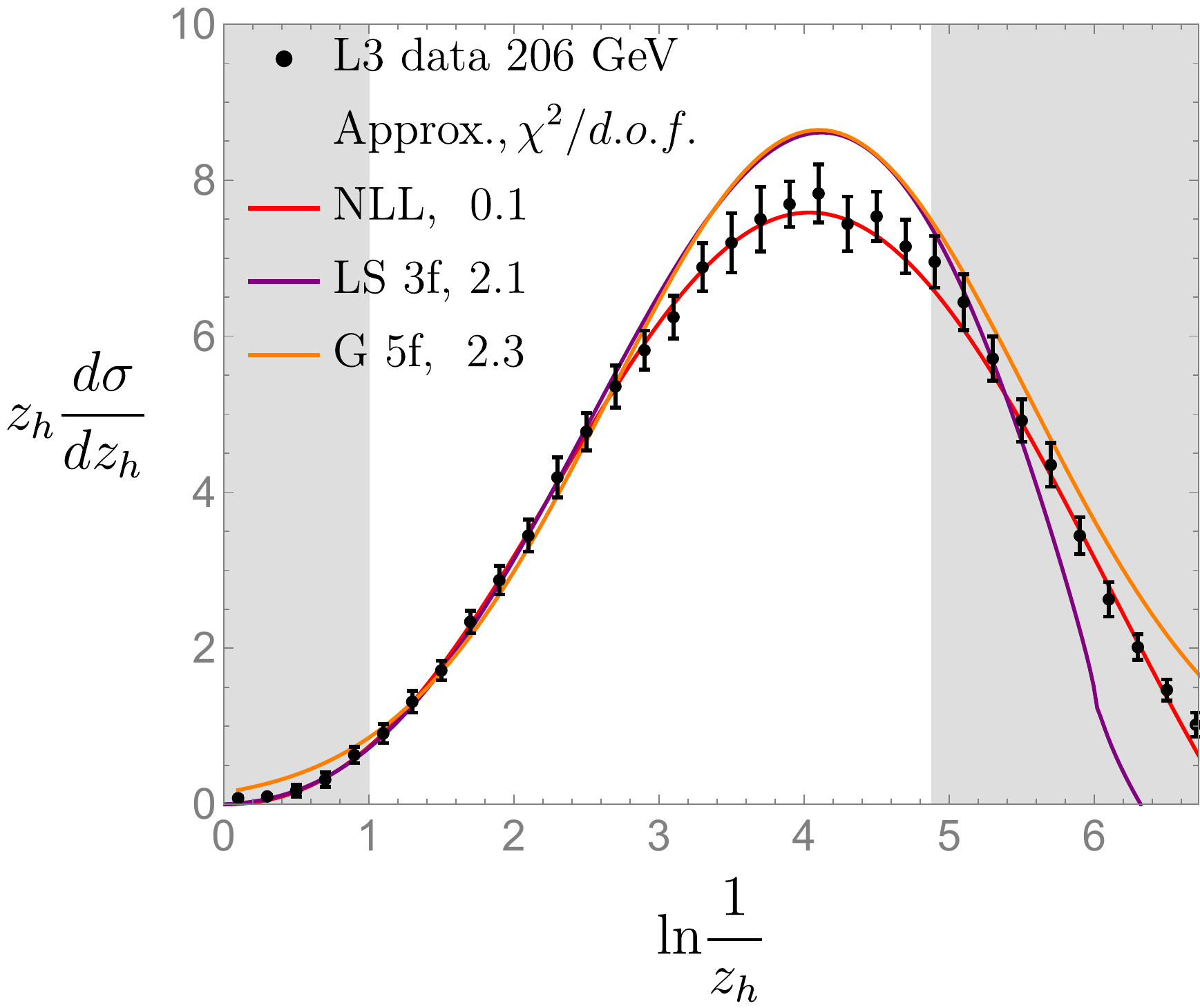}
   \caption{\label{fig:opal} The fragmentation spectrum for $e^+e^-\rightarrow h^{\pm}+X$ at center of mass energies $Q=91$ and $206$ GeV. The gray region is excluded from the \chidof determination based on hadron mass corrections or $\zh\sim O(1)$, and the quoted \chidof is the goodness-of-fit for that data set. Compared to the NLL curve is the evolved 3 flavor scheme MLLA limiting spectrum (purple LS) from ref. \cite{Akrawy:1990ha}, and a 5 flavor scheme evolving the moments of a Gaussian according to the MLLA anomalous dimension, as in ref. \cite{Fong:1990nt} (orange G). }
 \end{figure}

I can compare to the two common approximations for the MLLA, the so-called limiting spectrum and the Gaussian saddle-point. For completeness sake, I give the formula for the limiting spectrum and the Gaussian spectrum in app. \ref{sec:limit}, as well as the fit values. The spectra are supposed to be valid for $Q\rightarrow \infty$ and describe the peak region of the distribution, and good fits were obtained to Z-pole data in ref. \cite{Akrawy:1990ha} using 3 active quark flavors. In fig. \ref{fig:opal}, I present the original OPAL data, assuming a $1\%$ systematic error along with their quoted statistical uncertainty. I plot their MLLA limiting spectrum fit, as well as the Gaussian 5 active flavor fit, and quote \chidof for the fit region $1<-\text{ln}\zh<4.1$. Note that this fit region is smaller than the one used for the SLD data above, due to the Gaussian's more limited range of applicability. I see that all schemes appear capable of describing the peak of the OPAL data,\footnote{I also refit the NLL spectrum to the OPAL data, this results in a slightly larger normalization.} however, when evolved to the L3 scale, the predicted MLLA spectra fail to describe the data, except at moderate to small values of $-\text{ln}\zh$, while the NLL is uniformly valid throughout the fit region, and even beyond. I note also that the NLL result uses a physical number of active quark flavors at LEP energies.
 
\subsection{Low $Q$ Data} \label{sec:lowq}
 \begin{figure}\center
   \hspace{-10pt} \includegraphics[width=0.45\textwidth]{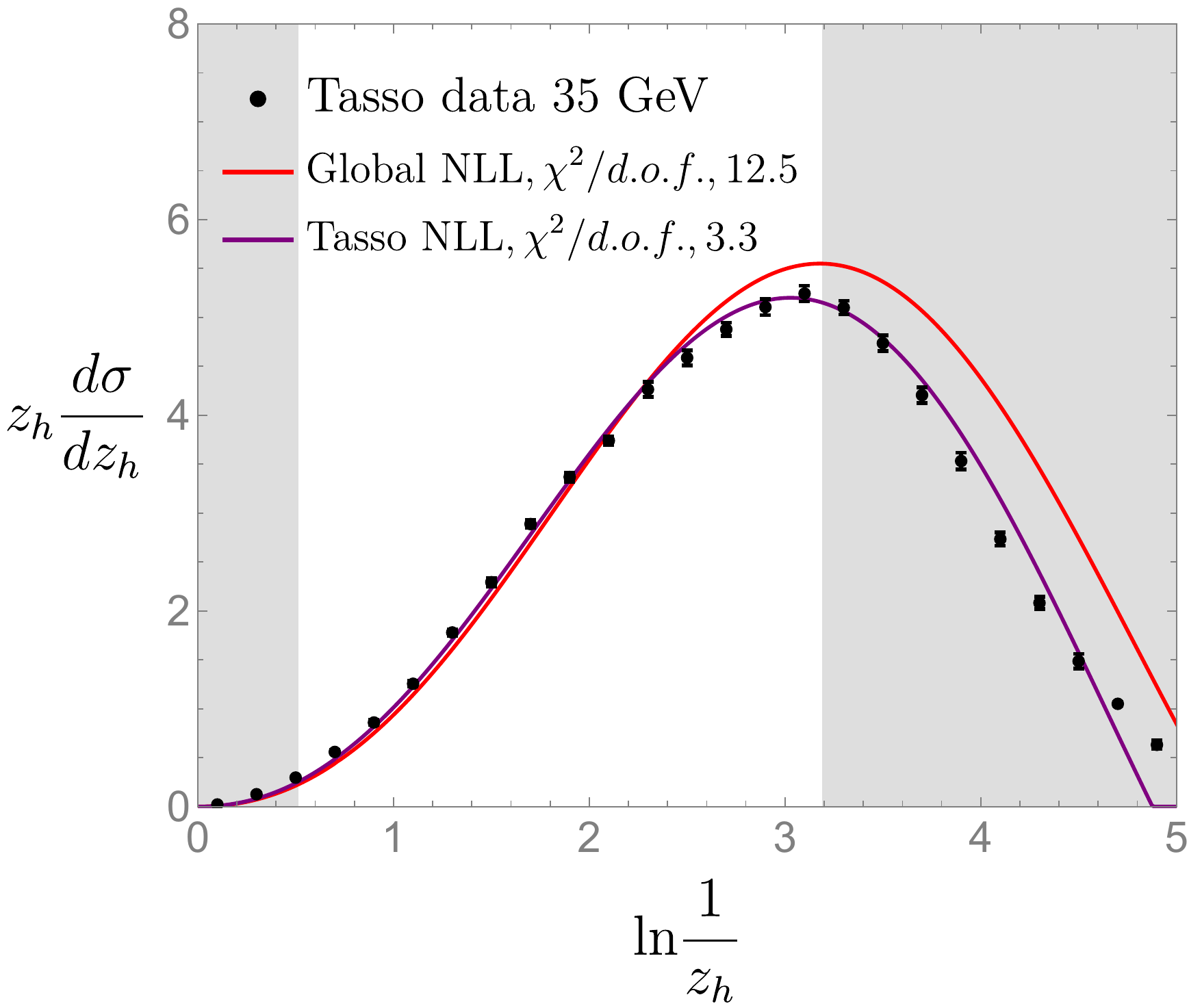}    \includegraphics[width=0.45\textwidth]{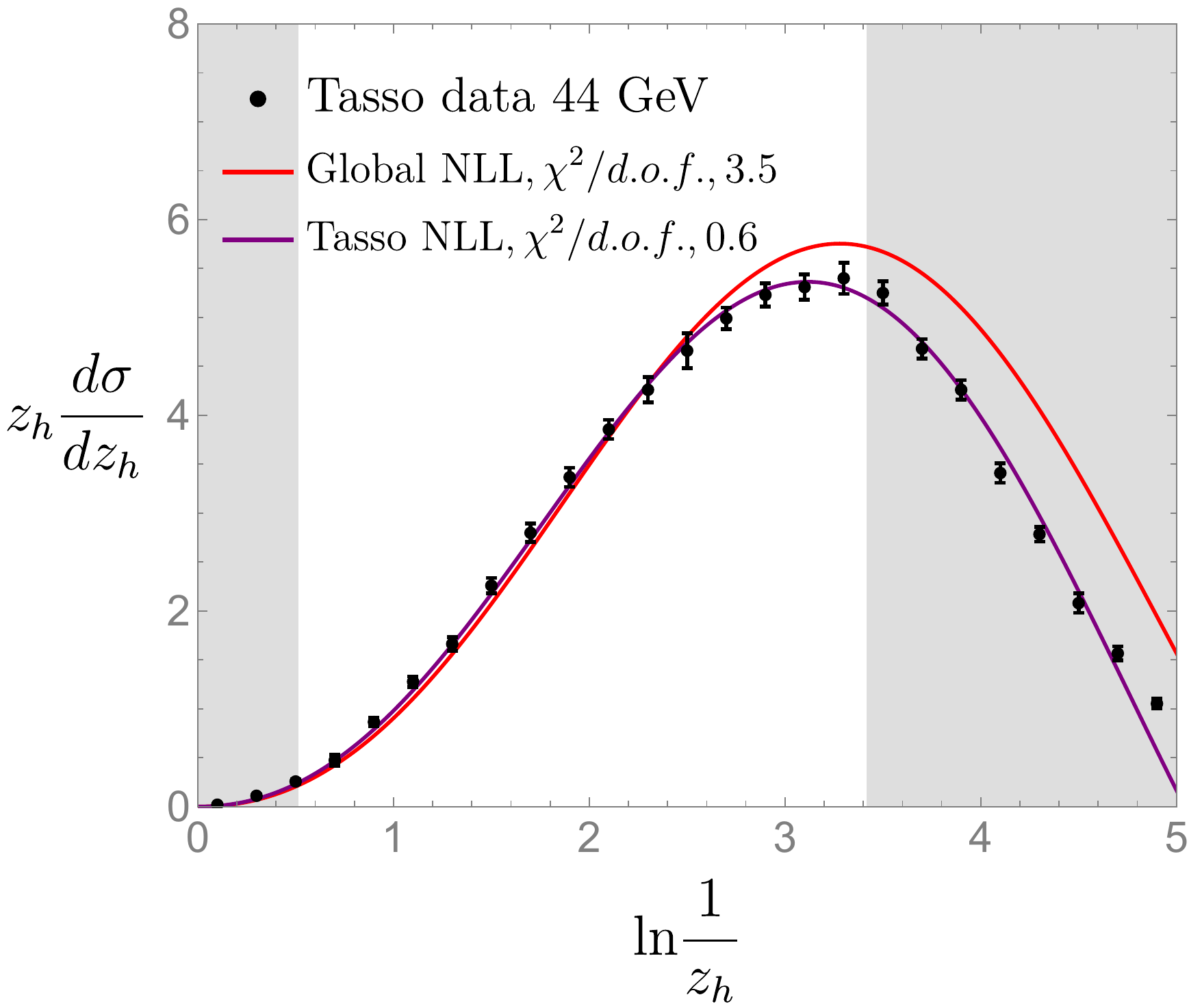}
   \caption{\label{fig:refit} The fragmentation spectrum for $e^+e^-\rightarrow h^{\pm}+X$ at center of mass energies $Q=35$ and $44$ GeV. The gray region is excluded from the \chidof determination based on hadron mass corrections or $\zh\sim O(1)$, and the quoted \chidof is the goodness-of-fit for that data set. I compare the global fit prediction to a fit to the lower Q data alone. The improved fit is achieved by only changing the scale at which splitting evolution is ended.}
 \end{figure}  
 The global fit described at the beginning of this section is dominated by the very precise SLD data at the Z-pole, which evolves nicely to describe the higher $Q$ LEP data. However it is at tension with the lower $Q$ TASSO data. I can refit to the lower $Q$ data, and in fig. \ref{fig:refit} I plot the new fit to the lower $Q$ data of TASSO, while plotting for comparison the global fit curve. The new fit differs only in that it prefers a higher value of $\Lambda$, at $2.0$ GeV. The normalization and coupling freezing scale are the same. That is, since $\Lambda_{\text{TASSO}}=2\Lambda_{\text{Global}}$, $c_{\text{TASSO}}=\frac{1}{4}c_{\text{Global}}$, so that the mass scale used to freeze the coupling remains the same,  $c_{\text{TASSO}} \Lambda_{\text{TASSO}}^2=c_{\text{Global}}\Lambda_{\text{Global}}^2$. I give the \chidof for both curves as a metric to see how the fit improves. This also nicely illustrates the impact of varying the end-point of just the evolution, as the other fit parameters are the same. Finally, I note I can drop the \chidof to $5.4$ for the TOPAZ data with a $\Lambda=1$ GeV like the global fit, but taking the normalization to $1.1$.

 At present, I can only speculate as to why ending the DGLAP evolution at a higher energy improves the fit. Clearly, there could be an issue of the missing higher order terms at NNLL or beyond in the resummation, which are becoming more important at lower vales of the hard initial scale. This should be investigated. However, the global fit and the specific TASSO fit are largely consistent at all $\zh$ values leading up to the peak, differing mainly in the peak itself and the region I omit due to potentially large hadron mass correction. Interestingly, the new fit interpolates into this region as well. This suggests that as hadron mass corrections become critical, they can be accounted for by tailoring the end-point of the DGLAP evolution to each hadron mass. Thus if I were to fit each charged species separately, varying the end point of evolution yet keeping the same basic fragmentation model proposed in this paper, I could by and large account for hadron mass effects. Then as $Q\rightarrow \infty$, the hadron mass effects become suppressed, and the endpoint of the DGLAP evolution tends to a universal value at the confinement scale. 

More concretely, to NLL accuracy, for each hadron species $h$, one would fit for a normalization $N_h$ and the endpoint of the evolution $\Lambda_h$ to eliminate hadron-mass corrections. This would result in the formula:
\begin{align}
   \fxsech_{h}\Big(\zh,Q^2,\Lambda^2\Big)&=\frac{N_{h}}{N(Q^2)}\int_{\zh}^{1}\frac{dz}{z} F\Big(z,Q^2,\Lambda^2_h\Big)\Bigg(U_{gg}\Big(\frac{\zh}{z},Q^2,\Lambda^2_h\Big)+ \frac{C_F}{C_A} U_{qg}\Big(\frac{\zh}{z},Q^2,\Lambda^2_h\Big)\Bigg)\,.
\end{align}  
The sum-rule could be enforced through the factor $N(Q^2)$ which would be determined from:
\begin{align}
1=\sum_h\int d\zh \zh \fxsech_{h}\Big(\zh,Q^2,\Lambda^2\Big)\,.
\end{align}  
Alternatively, one could enforce the sum-rule through matching the small-$\zh$ resummation to the large-$\zh$ terms of the perturbative expansion. The freezing of the coupling one could also vary with hadron species, or keep fixed for all species. Taking the latter option, this would increase the number of parameters for each hadron species for the fit to $2$, $\Lambda_h$, and  $N_h$, and the model for the coupling in the infra-red would be universal for all species. One could also allow the hadron to couple differently to quarks and gluons rather than use the ratio of color factors $C_F/C_A$. However, such investigations are beyond the scope of this paper.
 
 \subsection{Large $\zh$} \label{sec:large_zh}
 \begin{figure}\center
   \hspace{-10pt} \includegraphics[width=0.45\textwidth]{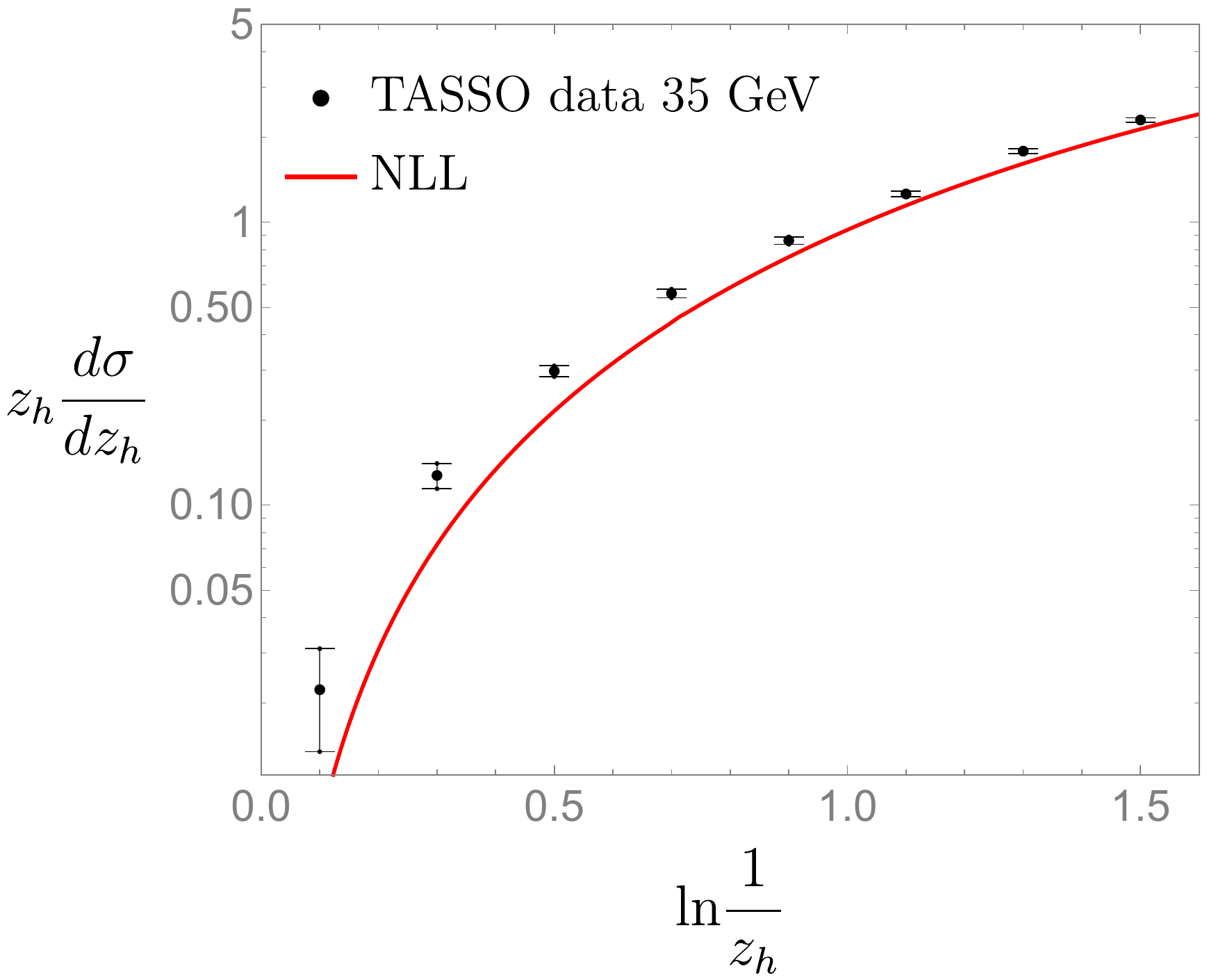}   \includegraphics[width=0.45\textwidth]{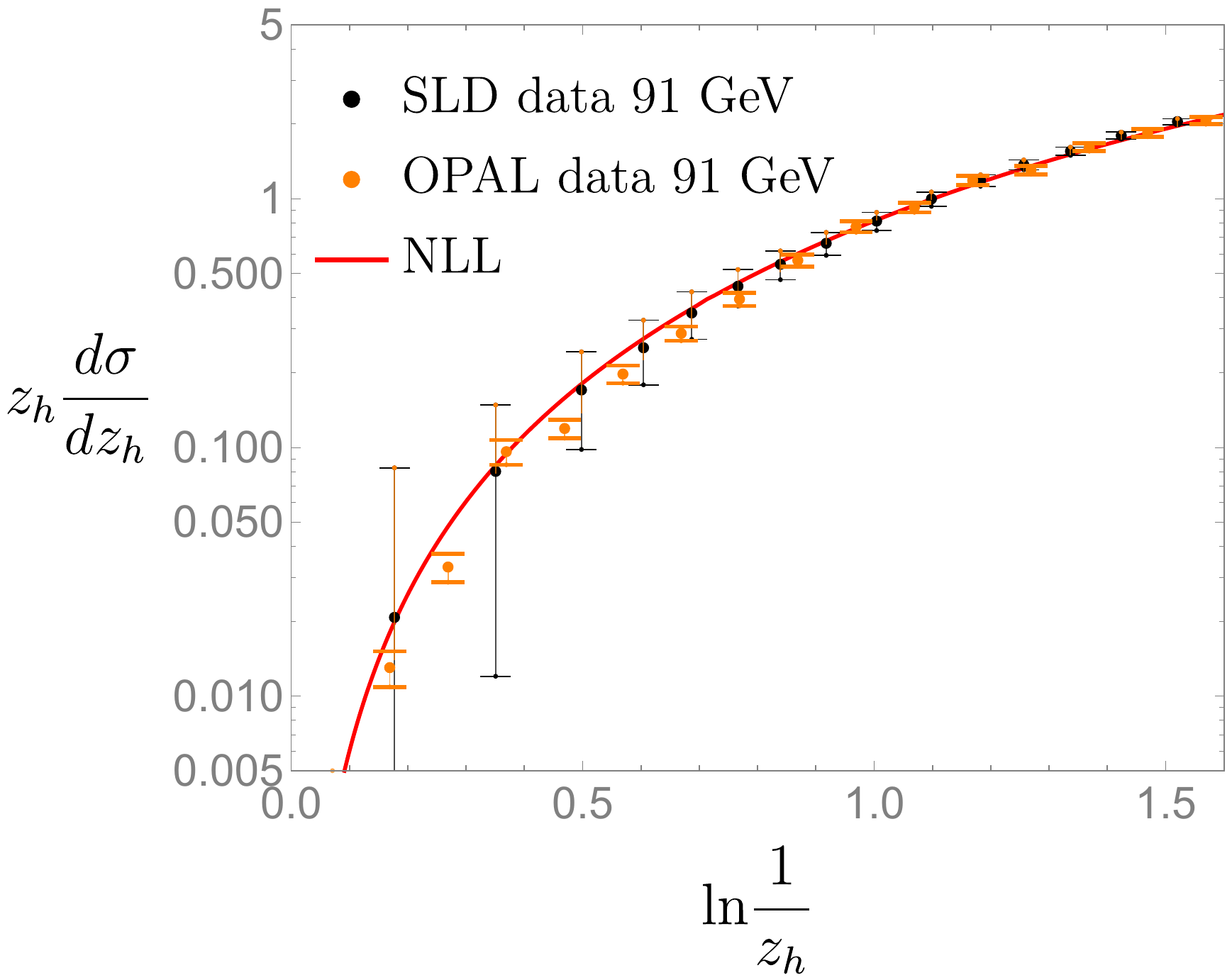} 
   \caption{\label{fig:largezh} The fragmentation spectrum extrapolated into the large $\zh$ region for $e^+e^-\rightarrow h^{\pm}+X$ at center of mass energies $Q=35$ and $91$. The region below ln$\frac{1}{\zh}<$0.5 was excluded from the fit.}
 \end{figure}  

\begin{figure}\center
   \includegraphics[width=0.45\textwidth]{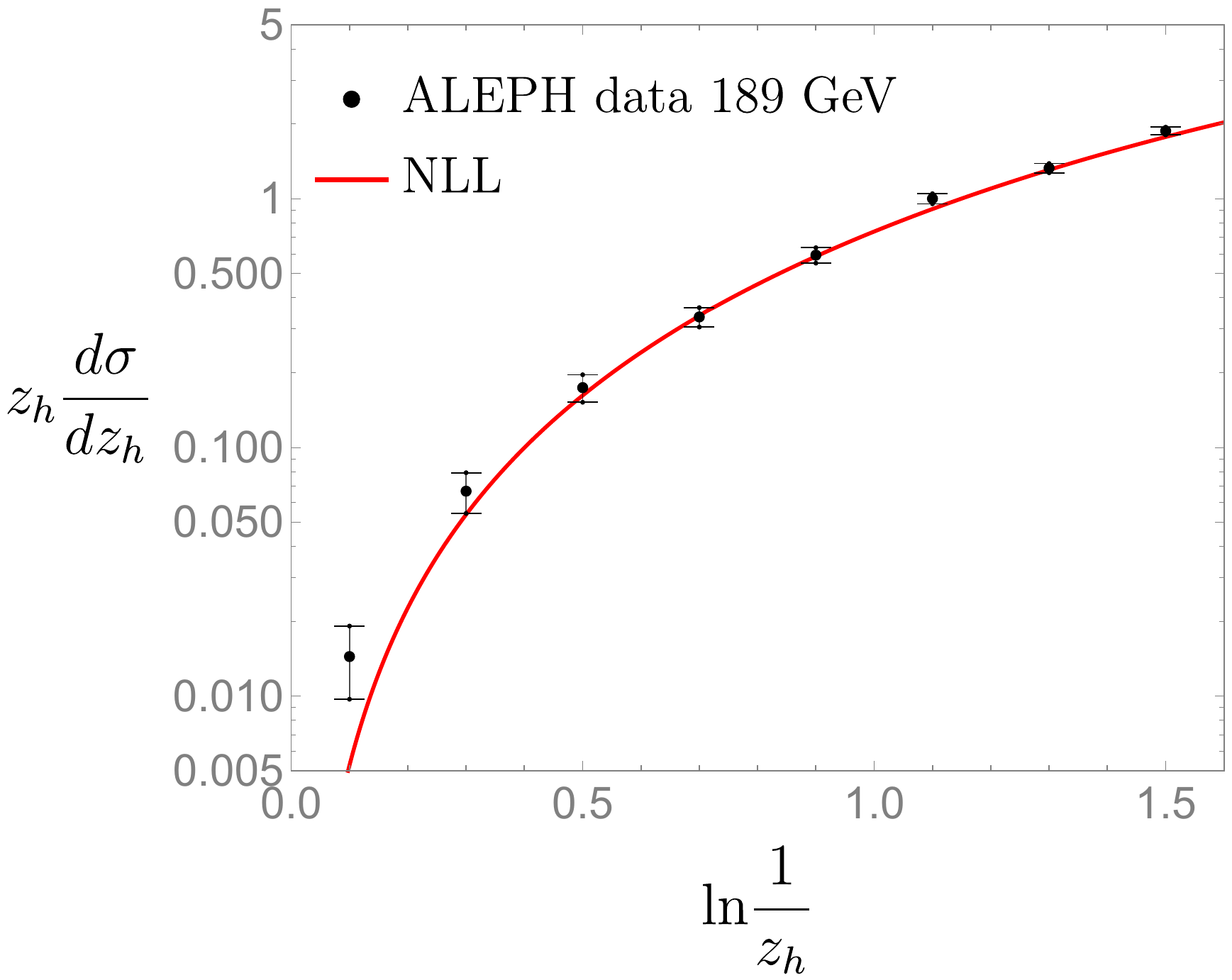}
   \caption{\label{fig:largezh_II} The fragmentation spectrum extrapolated into the large $\zh$ region for $e^+e^-\rightarrow h^{\pm}+X$ at center of mass energies $Q=189$. The region below ln$\frac{1}{\zh}<$0.5 was excluded from the fit.}
 \end{figure}

 I now present the extrapolation of the small-$\zh$ resummed cross-section into the large $\zh$ region. There is of course some ambiguity as to how to define the small $\zh$ region, but based on the comparison with the data, one appears to be solidly within this region as soon as $\zh\sim 0.3$, that is, ln$\frac{1}{\zh}\sim 1.2$. One would have the expectation that the cross-section should  \emph{not} be described at values of $\zh$ larger than this region by the small-$\zh$ resummation, but indeed from fig. \ref{fig:largezh} and  \ref{fig:largezh_II}, one finds that it \emph{can} interpolate almost to $\zh = 1$ for the high $Q^2$ SLD, OPAL, and ALEPH data, while also capturing the qualitative trend for the TASSO data. One would naively anticipate more severe deviations from the small-$\zh$ predictions within this region, and that matching to the $\zh\sim O(1)$ components of the hard function, jet function, and anomalous dimensions, as well as including threshold corrections would be required. While perhaps true for the TASSO data, the higher $Q^2$ data appears to not need as refined of a treatment. Investigation into the origins of this rather unexpected validity of the small-$\zh$ resummations to the whole physical region is beyond the scope of this paper.

 \section{Conclusions}

 The concept of local parton-hadron-duality gives important insight into the process of hadronization, and helps explain the success of QCD perturbation theory, beyond the traditional estimates of QCD factorization. It is important to note that using the traditional power counting regions for the modes informing the factorization for fragmentation, local parton-hadron-duality could be considered surprising, see refs. \cite{Collins:2016ztc,Collins:2018teg}. And indeed, treating the fragmentation spectrum with the standard momentum regions, tacitly or explicitly assuming $\zh\sim O(1)$, leads to using the fixed-order expansion for all quantities. This does not achieve the fragmentation spectrum given here. A soft resummation is needed.

 However, the original formulation of LPHD rested on unsystematic approximations to the QCD perturbation theory, making it unclear whether the success of the LPHD hypothesis rested on these unsystematic approximations, and whether a more grounded approach to QCD perturbation theory would destroy the LPHD concept. I find that this is not the case: systematically improving the small-$\zh$ resummation to NLL accuracy, in the coefficient function, anomalous dimension, \emph{and} jet/fragmentation function gives a uniformly valid description of the fragmentation data, by extrapolating semi-inclusive jet production to hadrons and freezing the infra-red coupling at a quasi-perturbative scale. Even though the small-$\zh$ resummed ansatz worked well, there is still room for improvement. First one can extend the small-$\zh$ resummation to NNLL and even N$^3$LL order. Further, one can match to the threshold region at $\zh = 1$ (see refs. \cite{Dai:2017dpc,Liu:2018ktv}), hopefully enforcing the sum-rule for the spectrum ``naturally.''  Moreover, one would also like to use the small-$\zh$ resummation to improve the description of jet production rather than just hadron fragmentation, constructing a jet function that is uniformly valid in all regions of the energy fraction.

From a more formal point of view, the results are very pleasing: the fragmentation spectrum of hadrons is truly determined by reciprocity with the space-like branching process, when expressed in angular ordered evolution equations. Initially this reciprocity was postulated for the anomalous dimensions alone in refs. \cite{Dokshitzer:2005bf,Basso:2006nk}, and found a critical role in the scaling properties of energy-correlation functions (refs. \cite{Dixon:2019uzg,Chen:2020vvp}). But to capture the fragmentation spectrum, this reciprocity must be extended to the full cross-section as done here.

\section{Acknowledgments}
D.N. wishes to thank Felix Ringer and Nobou Sato for discussion of fragmentation and Wouter Waalewijn for reading the manuscript. D.N. was supported by the Department of Energy under Contract DE-AC52-06NA25396 at LANL and through the LANL/LDRD Program.

\appendix

\section{MLLA spectra}\label{sec:limit}
The limiting spectrum of the MLLA is given as (refs. \cite{Dokshitzer:1991ej,Akrawy:1990ha}):
\begin{align}
  \zh \fxsech(\zh,Q^2,\Lambda^2)&=\kappa\frac{4N_c}{b}\Gamma(B)\int\displaylimits_{-\frac{\pi}{2}}^{{\frac{\pi}{2}}}\frac{d\tau}{\pi}e^{-B\alpha(\tau)}\Bigg(\frac{C(\alpha(\tau),L,Y)}{\frac{4C_A}{b}Y\frac{\alpha(\tau)}{\sinh\alpha(\tau)}}\Bigg)^{B/2}I_{B}\Bigg(\sqrt{A(\alpha(\tau),L,Y)}\Bigg)\,\\
  A(\alpha,L,Y)&=\frac{16N_c}{b}Y\frac{\alpha}{\sinh\alpha}C(\alpha,L,Y),\nonumber\\
  C(\alpha,L,Y)&=\cosh\alpha+\Big(\frac{2L}{Y}-1\Big)\sinh \alpha, \nonumber\\
  Y&=\text{ln}\frac{Q}{2\Lambda},\quad L=\text{ln}\frac{1}{\zh},\nonumber\\
  \alpha(\tau)&=\alpha_0+i\tau,\quad \tanh\alpha_0 = \Big(1-\frac{2L}{Y}\Big),\nonumber\\
  B&=\frac{1}{b}\Big(\frac{11N_c}{3}+\frac{2n_f}{3N_c^2}\Big),\quad b=\frac{11N_c}{3}-\frac{2n_f}{3}.
\end{align}
$N_c$ is the number of colors, $n_f$ the number of active quarks, $I_B$ is the modified Bessel function. One fits for the normalization $\kappa$ and confinement scale $\Lambda$. For the 3 flavor OPAL fit, I have $\kappa=1.28$, and $\Lambda=0.253$ GeV. 

The Gaussian approximation is given as:
\begin{align}
  \zh \fxsech(\zh,Q^2,\Lambda^2)&=\kappa N(Q^2,\Lambda^2) \Bigg(\frac{c_1}{Y^{3/2}\pi}\Bigg)^{1/2}\text{exp}\Bigg(-\frac{c_1}{Y^{3/2}}\Big(L-\frac{1}{2}Y-c_2\sqrt{Y}+o\Big)^2\Bigg)\,,\\
  c_1&=\sqrt{36 N_c/b}\,,\\
  c_2&=B\sqrt{b/(16 N_c)}\,,\\
  N(Q^2,\Lambda^2)&=\text{exp}\Bigg(4\sqrt{\frac{N_cY}{b}}-\Big(B-\frac{1}{2}\Big)\text{ln}\sqrt{Y}\Bigg)\,.
\end{align}
$L,B,Y$ and $b$ are defined as above. I have for the fit to OPAL data: $\Lambda=0.2$ GeV, $\kappa=0.148$, and $o=0.5$ for $n_f=5$.

\section{Implementing DGLAP}\label{sec:numerics_notes}
I evolve the DGLAP equations in momentum space, using a time variable;
\begin{align}
t=\text{ln}\frac{\mu^2_i}{\mu^2_f}, \mu_i>\mu_f\,.
\end{align}  
The current mass scale of the evolution is given by $\mu_ie^{-t}$, and thus this is the scale at which $\alpha_s$ would be evaluated at a given evolution time. I am evolving from the high scale down to a low scale I wish to solve the equation:
\begin{align}\label{eq:evo_eq}
\frac{d}{dt}H_{i}\Big(x,t\Big)&=\sum_{j}\int_{x}^{1}\frac{dz}{z}P_{ij}\Big(\frac{x}{z},t\Big)H_{j}\Big(z,t\Big)\,.
\end{align}
$i,j$ are flavor indices. To do so I discretize in space and time:
\begin{align}
  &x_{min}<x_1<x_2<...<x_{N_{max}}<1\,,\\
  \Delta x_{\alpha}&=x_{\alpha}-x_{\alpha-1}\,,\\
  t_{n+1}&=t_{n}+\Delta t_{n+1}\,.
\end{align}  
The code can accept arbitrary gridding in momentum space and the time variable, so that the time step taken depends on which time I am stepping to. This allows for efficient evolution at both large and small times, depending on how fast $\alpha_s$ is changing. I use the fourth order Runge-Kutta solution to eq. \eqref{eq:evo_eq}:
\begin{align}
  K_{i}^{1}(x_\alpha)&=\sum_{j}\sum_{\beta=\alpha}^{N_{max}}\Delta t \frac{\Delta x_{\beta}}{x_\beta}P_{ij}\Big(\frac{x_\alpha}{x_{\beta}},t_n\Big)H_{j}\Big(x_{\beta},t_{n}\Big)\,,\\
  K_{i}^{2}(x_\alpha)&=\sum_{j}\sum_{\beta=\alpha}^{N_{max}}\Delta t \frac{\Delta x_{\beta}}{x_\beta}P_{ij}\Big(\frac{x_\alpha}{x_{\beta}},t_n+\frac{\Delta t}{2}\Big)\Bigg[H_{j}\Big(x_{\beta},t_{n}\Big)+\frac{1}{2}K_{j}^{1}(x_{\beta})\Bigg]\,,\\
  K_{i}^{3}(x_\alpha)&=\sum_{j}\sum_{\beta=\alpha}^{N_{max}}\Delta t \frac{\Delta x_{\beta}}{x_\beta}P_{ij}\Big(\frac{x_\alpha}{x_{\beta}},t_n+\frac{\Delta t}{2}\Big)\Bigg[H_{j}\Big(x_{\beta},t_{n}\Big)+\frac{1}{2}K_{j}^{2}(x_{\beta})\Bigg]\,,\\
  K_{i}^{4}(x_\alpha)&=\sum_{j}\sum_{\beta=\alpha}^{N_{max}}\Delta t \frac{\Delta x_{\beta}}{x_\beta}P_{ij}\Big(\frac{x_\alpha}{x_{\beta}},t_n+\Delta t\Big)\Bigg[H_{j}\Big(x_{\beta},t_{n}\Big)+K_{j}^{3}(x_{\beta})\Bigg]\,,\\
  H_{i}\Big(x_{\alpha},t_{n+1}\Big)&=H_{i}\Big(x_{\alpha},t_{n}\Big)+\frac{1}{6}\Big(K_{i}^{1}(x_\alpha)+2K_{i}^{2}(x_\alpha)+2K_{i}^{3}(x_\alpha)+K_{i}^{4}(x_\alpha)\Big)\,.
\end{align}
Note that the above numerical evaluations of the convolution integral assume no plus distributions in $P_{ij}$. Terms with a plus distribution in the splitting function $P_{ij}(z)$ must be handled more carefully. In particular, I use the formula:
\begin{align}
  \int_{x}^{1}\frac{dz}{z}\Big[g\Big(\frac{x}{z}\Big)\Big]_+ f(z)&=\int_{x}^{1}\frac{dz}{z}g\Big(\frac{x}{z}\Big)\Big( f(z)-\frac{x}{z}f(x)\Big)-f(x)\int_0^{x}dz g(z)\,,\\
  &=\sum_{\beta=(\alpha+1)}^{N_{max}} \frac{\Delta x_{\beta}}{x_\beta}g\Big(\frac{x_\alpha}{x_{\beta}}\Big)\Bigg[f(x_{\beta})-\frac{x_{\alpha}}{x_{\beta}}f(x_{\alpha})\Bigg]-f(x_{\alpha})\sum_{\beta=min}^{\alpha} \Delta x_{\beta} g(x_\beta)\,.
\end{align}  
Note that I drop the bin that explicitly contains the singularity. This does not pose a problem, since the plus distribution regulates this singularity. That is to say, at $x_{\beta}\rightarrow x_{\alpha}$, I have a $0/0\sim finite$, which then multiplies the bin width. Thus the contribution of this bin to the integral is of the order the bin width, and in the limit that the bin width goes to zero, this contribution vanishes.

Thus I have two convolution routines, one that computes the contribution from the plussed parts of the splitting function, and the other from the regular parts (including any delta function contributions). The delta functions within the splitting functions are represented as:
\begin{align}
\delta(1-x)=\frac{1}{\Delta x_\beta}\Theta\Big(1-x_{\beta}\Big)\Theta\Big(\Delta x_\beta-(1-x_\beta)\Big)\,,
\end{align}
where $\Delta x_\beta$ is the width of the bin containing $x_\beta$, and $x\approx x_{\beta}$. However, for the delta function initial condition to the evolution ($H(x)=\delta(1-x)$), I instead use the formula:
\begin{align}
H(x_\beta)&=\frac{2}{\Delta x_{N_{max}}\sqrt{2\pi}}\text{exp}\Big(-\frac{(1-x_\beta)^2}{2\Delta x_{N_{max}}^2}\Big)\,.
\end{align}

While for this paper I evolve only the small-$\zh$ resummed splitting functions, so I need not deal with plus functions, I have tested that the code agrees with mellin space evolution for full QCD if I use the standard leading order DGLAP evolution kernels instead. If I ignore running coupling effects, I can also solve the evolution in pure Yang-Mills theory in the small-$\zh$ regime analytically to LL accuracy. Again I check the code reproduces these results.

\bibliographystyle{JHEP}
\bibliography{bibliography}

\providecommand{\href}[2]{#2}\begingroup\raggedright\begin{thebibliography}{10}

\bibitem{Azimov:1984np}
Y.~I. Azimov, Y.~L. Dokshitzer, V.~A. Khoze, and S.~Troyan, {\it {Similarity of
  Parton and Hadron Spectra in QCD Jets}},  {\em Z. Phys. C} {\bf 27} (1985)
  65--72.

\bibitem{Khoze:1996dn}
V.~A. Khoze and W.~Ochs, {\it {Perturbative QCD approach to multiparticle
  production}},  {\em Int. J. Mod. Phys.} {\bf A12} (1997) 2949--3120,
  [\href{http://arxiv.org/abs/hep-ph/9701421}{{\tt hep-ph/9701421}}].

\bibitem{Metz:2016swz}
A.~Metz and A.~Vossen, {\it {Parton Fragmentation Functions}},  {\em Prog.
  Part. Nucl. Phys.} {\bf 91} (2016) 136--202,
  [\href{http://arxiv.org/abs/1607.02521}{{\tt arXiv:1607.02521}}].

\bibitem{Fong:1989qy}
C.~Fong and B.~Webber, {\it {Higher Order {QCD} Corrections to Hadron Energy
  Distributions in Jets}},  {\em Phys. Lett. B} {\bf 229} (1989) 289--292.

\bibitem{Fong:1990nt}
C.~Fong and B.~Webber, {\it {One and two particle distributions at small x in
  QCD jets}},  {\em Nucl. Phys. B} {\bf 355} (1991) 54--81.

\bibitem{Albino:2004yg}
S.~Albino, B.~Kniehl, and G.~Kramer, {\it {Low x particle spectra in the
  Modified Leading Logarithm Approximation}},  {\em Eur. Phys. J. C} {\bf 38}
  (2004) 177--185, [\href{http://arxiv.org/abs/hep-ph/0408112}{{\tt
  hep-ph/0408112}}].

\bibitem{Dokshitzer:1991ej}
Y.~L. Dokshitzer, V.~A. Khoze, and S.~Troian, {\it {Inclusive particle spectra
  from QCD cascades}},  {\em Int. J. Mod. Phys. A} {\bf 7} (1992) 1875--1906.

\bibitem{Akrawy:1990ha}
{\bf OPAL} Collaboration, M.~Akrawy et~al., {\it {A Study of coherence of soft
  gluons in hadron jets}},  {\em Phys. Lett. B} {\bf 247} (1990) 617--628.

\bibitem{Perez-Ramos:2013eba}
R.~Perez-Ramos and D.~d'Enterria, {\it {Energy evolution of the moments of the
  hadron distribution in QCD jets including NNLL resummation and NLO
  running-coupling corrections}},  {\em JHEP} {\bf 08} (2014) 068,
  [\href{http://arxiv.org/abs/1310.8534}{{\tt arXiv:1310.8534}}].

\bibitem{Vogt:2011jv}
A.~Vogt, {\it {Resummation of small-x double logarithms in QCD: semi-inclusive
  electron-positron annihilation}},  {\em JHEP} {\bf 10} (2011) 025,
  [\href{http://arxiv.org/abs/1108.2993}{{\tt arXiv:1108.2993}}].

\bibitem{Kom:2012hd}
C.~H. Kom, A.~Vogt, and K.~Yeats, {\it {Resummed small-x and first-moment
  evolution of fragmentation functions in perturbative QCD}},  {\em JHEP} {\bf
  10} (2012) 033, [\href{http://arxiv.org/abs/1207.5631}{{\tt
  arXiv:1207.5631}}].

\bibitem{Neill:2020bwv}
D.~Neill and F.~Ringer, {\it {Soft Fragmentation on the Celestial Sphere}},
  {\em JHEP} {\bf 06} (2020) 086, [\href{http://arxiv.org/abs/2003.02275}{{\tt
  arXiv:2003.02275}}].

\bibitem{Dasgupta:2014yra}
M.~Dasgupta, F.~Dreyer, G.~P. Salam, and G.~Soyez, {\it {Small-radius jets to
  all orders in QCD}},  {\em JHEP} {\bf 04} (2015) 039,
  [\href{http://arxiv.org/abs/1411.5182}{{\tt arXiv:1411.5182}}].

\bibitem{Dasgupta:2016bnd}
M.~Dasgupta, F.~A. Dreyer, G.~P. Salam, and G.~Soyez, {\it {Inclusive jet
  spectrum for small-radius jets}},  {\em JHEP} {\bf 06} (2016) 057,
  [\href{http://arxiv.org/abs/1602.01110}{{\tt arXiv:1602.01110}}].

\bibitem{Kang:2016mcy}
Z.-B. Kang, F.~Ringer, and I.~Vitev, {\it {The semi-inclusive jet function in
  SCET and small radius resummation for inclusive jet production}},  {\em JHEP}
  {\bf 10} (2016) 125, [\href{http://arxiv.org/abs/1606.06732}{{\tt
  arXiv:1606.06732}}].

\bibitem{Dai:2016hzf}
L.~Dai, C.~Kim, and A.~K. Leibovich, {\it {Fragmentation of a Jet with Small
  Radius}},  {\em Phys. Rev.} {\bf D94} (2016), no.~11 114023,
  [\href{http://arxiv.org/abs/1606.07411}{{\tt arXiv:1606.07411}}].

\bibitem{Kang:2017mda}
Z.-B. Kang, F.~Ringer, and W.~J. Waalewijn, {\it {The Energy Distribution of
  Subjets and the Jet Shape}},  {\em JHEP} {\bf 07} (2017) 064,
  [\href{http://arxiv.org/abs/1705.05375}{{\tt arXiv:1705.05375}}].

\bibitem{Dokshitzer:1995qm}
Y.~L. Dokshitzer, G.~Marchesini, and B.~Webber, {\it {Dispersive approach to
  power behaved contributions in QCD hard processes}},  {\em Nucl. Phys. B}
  {\bf 469} (1996) 93--142, [\href{http://arxiv.org/abs/hep-ph/9512336}{{\tt
  hep-ph/9512336}}].

\bibitem{Mattingly:1993ej}
A.~Mattingly and P.~M. Stevenson, {\it {Optimization of R(e+ e-) and 'freezing'
  of the QCD couplant at low-energies}},  {\em Phys. Rev. D} {\bf 49} (1994)
  437--450, [\href{http://arxiv.org/abs/hep-ph/9307266}{{\tt hep-ph/9307266}}].

\bibitem{Anderle:2016czy}
D.~P. Anderle, T.~Kaufmann, M.~Stratmann, and F.~Ringer, {\it {Fragmentation
  Functions Beyond Fixed Order Accuracy}},  {\em Phys. Rev.} {\bf D95} (2017),
  no.~5 054003, [\href{http://arxiv.org/abs/1611.03371}{{\tt
  arXiv:1611.03371}}].

\bibitem{Albino:2011si}
S.~Albino, P.~Bolzoni, B.~A. Kniehl, and A.~Kotikov, {\it {Fully
  double-logarithm-resummed cross sections}},  {\em Nucl. Phys.} {\bf B851}
  (2011) 86--103, [\href{http://arxiv.org/abs/1104.3018}{{\tt
  arXiv:1104.3018}}].

\bibitem{Collins:1989gx}
J.~C. Collins, D.~E. Soper, and G.~F. Sterman, {\em {Factorization of Hard
  Processes in QCD}}, vol.~5, pp.~1--91.
\newblock 1989.
\newblock \href{http://arxiv.org/abs/hep-ph/0409313}{{\tt hep-ph/0409313}}.

\bibitem{Chen:2020uvt}
H.~Chen, T.-Z. Yang, H.~X. Zhu, and Y.~J. Zhu, {\it {Analytic Continuation and
  Reciprocity Relation for Collinear Splitting in QCD}},
  \href{http://arxiv.org/abs/2006.10534}{{\tt arXiv:2006.10534}}.

\bibitem{Moch:2007tx}
S.~Moch and A.~Vogt, {\it {On third-order timelike splitting functions and
  top-mediated Higgs decay into hadrons}},  {\em Phys. Lett.} {\bf B659} (2008)
  290--296, [\href{http://arxiv.org/abs/0709.3899}{{\tt arXiv:0709.3899}}].

\bibitem{Gituliar:2015pra}
O.~Gituliar and S.~Moch, {\it {Towards three-loop QCD corrections to the
  time-like splitting functions}},  {\em Acta Phys. Polon.} {\bf B46} (2015),
  no.~7 1279--1289, [\href{http://arxiv.org/abs/1505.02901}{{\tt
  arXiv:1505.02901}}].

\bibitem{Ciafaloni:2005cg}
M.~Ciafaloni and D.~Colferai, {\it {Dimensional regularisation and
  factorisation schemes in the BFKL equation at subleading level}},  {\em JHEP}
  {\bf 09} (2005) 069, [\href{http://arxiv.org/abs/hep-ph/0507106}{{\tt
  hep-ph/0507106}}].

\bibitem{Ciafaloni:2006yk}
M.~Ciafaloni, D.~Colferai, G.~P. Salam, and A.~M. Stasto, {\it {Minimal
  subtraction vs. physical factorisation schemes in small-x QCD}},  {\em Phys.
  Lett.} {\bf B635} (2006) 320--329,
  [\href{http://arxiv.org/abs/hep-ph/0601200}{{\tt hep-ph/0601200}}].

\bibitem{Beneke:1997zp}
M.~Beneke and V.~A. Smirnov, {\it {Asymptotic expansion of Feynman integrals
  near threshold}},  {\em Nucl. Phys. B} {\bf 522} (1998) 321--344,
  [\href{http://arxiv.org/abs/hep-ph/9711391}{{\tt hep-ph/9711391}}].

\bibitem{Jantzen:2011nz}
B.~Jantzen, {\it {Foundation and generalization of the expansion by regions}},
  {\em JHEP} {\bf 12} (2011) 076, [\href{http://arxiv.org/abs/1111.2589}{{\tt
  arXiv:1111.2589}}].

\bibitem{Braunschweig:1990yd}
{\bf TASSO} Collaboration, W.~Braunschweig et~al., {\it {Global Jet Properties
  at 14-{GeV} to 44-{GeV} Center-of-mass Energy in $e^+ e^-$ Annihilation}},
  {\em Z. Phys. C} {\bf 47} (1990) 187--198.

\bibitem{Itoh:1994kb}
{\bf TOPAZ} Collaboration, R.~Itoh et~al., {\it {Measurement of inclusive
  particle spectra and test of MLLA prediction in e+ e- annihilation at
  s**(1/2) = 58-GeV}},  {\em Phys. Lett. B} {\bf 345} (1995) 335--342,
  [\href{http://arxiv.org/abs/hep-ex/9412015}{{\tt hep-ex/9412015}}].

\bibitem{Abe:2003iy}
{\bf SLD} Collaboration, K.~Abe et~al., {\it {Production of pi+, pi-, K+, K-, p
  and anti-p in light (uds), c and b jets from Z0 decays}},  {\em Phys. Rev. D}
  {\bf 69} (2004) 072003, [\href{http://arxiv.org/abs/hep-ex/0310017}{{\tt
  hep-ex/0310017}}].

\bibitem{Heister:2003aj}
{\bf ALEPH} Collaboration, A.~Heister et~al., {\it {Studies of QCD at e+ e-
  centre-of-mass energies between 91-GeV and 209-GeV}},  {\em Eur. Phys. J. C}
  {\bf 35} (2004) 457--486.

\bibitem{Achard:2004sv}
{\bf L3} Collaboration, P.~Achard et~al., {\it {Studies of hadronic event
  structure in $e^{+} e^{-}$ annihilation from 30-GeV to 209-GeV with the L3
  detector}},  {\em Phys. Rept.} {\bf 399} (2004) 71--174,
  [\href{http://arxiv.org/abs/hep-ex/0406049}{{\tt hep-ex/0406049}}].

\bibitem{Nason:1993xx}
P.~Nason and B.~Webber, {\it {Scaling violation in e+ e- fragmentation
  functions: QCD evolution, hadronization and heavy quark mass effects}},  {\em
  Nucl. Phys. B} {\bf 421} (1994) 473--517. [Erratum: Nucl.Phys.B 480, 755
  (1996)].

\bibitem{Collins:2016ztc}
J.~Collins, {\it {Do fragmentation functions in factorization theorems
  correctly treat non-perturbative effects?}},  {\em PoS} {\bf QCDEV2016}
  (2017) 003, [\href{http://arxiv.org/abs/1610.09994}{{\tt arXiv:1610.09994}}].

\bibitem{Collins:2018teg}
J.~Collins and T.~C. Rogers, {\it {Graphical Structure of Hadronization and
  Factorization in Hard Collisions}},
  \href{http://arxiv.org/abs/1801.02704}{{\tt arXiv:1801.02704}}.

\bibitem{Dai:2017dpc}
L.~Dai, C.~Kim, and A.~K. Leibovich, {\it {Fragmentation to a jet in the large
  $z$ limit}},  {\em Phys. Rev. D} {\bf 95} (2017), no.~7 074003,
  [\href{http://arxiv.org/abs/1701.05660}{{\tt arXiv:1701.05660}}].

\bibitem{Liu:2018ktv}
X.~Liu, S.-O. Moch, and F.~Ringer, {\it {Phenomenology of single-inclusive jet
  production with jet radius and threshold resummation}},  {\em Phys. Rev. D}
  {\bf 97} (2018), no.~5 056026, [\href{http://arxiv.org/abs/1801.07284}{{\tt
  arXiv:1801.07284}}].

\bibitem{Dokshitzer:2005bf}
{\relax Yu}.~L. Dokshitzer, G.~Marchesini, and G.~P. Salam, {\it {Revisiting
  parton evolution and the large-x limit}},  {\em Phys. Lett.} {\bf B634}
  (2006) 504--507, [\href{http://arxiv.org/abs/hep-ph/0511302}{{\tt
  hep-ph/0511302}}].

\bibitem{Basso:2006nk}
B.~Basso and G.~P. Korchemsky, {\it {Anomalous dimensions of high-spin
  operators beyond the leading order}},  {\em Nucl. Phys.} {\bf B775} (2007)
  1--30, [\href{http://arxiv.org/abs/hep-th/0612247}{{\tt hep-th/0612247}}].

\bibitem{Dixon:2019uzg}
L.~J. Dixon, I.~Moult, and H.~X. Zhu, {\it {Collinear limit of the
  energy-energy correlator}},  {\em Phys. Rev. D} {\bf 100} (2019), no.~1
  014009, [\href{http://arxiv.org/abs/1905.01310}{{\tt arXiv:1905.01310}}].

\bibitem{Chen:2020vvp}
H.~Chen, I.~Moult, X.~Zhang, and H.~X. Zhu, {\it {Rethinking Jets with Energy
  Correlators: Tracks, Resummation and Analytic Continuation}},  {\em Phys.
  Rev. D} {\bf 102} (2020), no.~5 054012,
  [\href{http://arxiv.org/abs/2004.11381}{{\tt arXiv:2004.11381}}].

\end{thebibliography}\endgroup


\end{document}